\documentclass[usenatbib,useAMS]{mn2e}

\usepackage{color}
\usepackage{times}
\usepackage{epsf}
\usepackage[dvips]{epsfig}

\begin{document}

\label{firstpage}

\title[Formation of the thick disc of the Milky Way]
{Popping star clusters as building blocks of the Milky Way Thick
  Disc} 

\author[Assmann et al.]{
  P. Assmann$^{1}$ \thanks{E-mail: passmann@astro-udec.cl},
  M. Fellhauer$^{1}$ \thanks{mfellhauer@astro-udec.cl},
  P. Kroupa$^{2}$ \thanks{pavel@astro.uni-bonn.de},
  R.C. Br\"{u}ns$^{2}$ \thanks{rcbruens@astro.uni-bonn.de},
  R. Smith$^{1}$ \thanks{rsmith@astro-udec.cl}\\
  $^{1}$ Departamento de Astronom\'{i}a, Universidad de
  Concepci\'{o}n,
  Casila 160-C, Concepci\'{o}n, Chile \\
  $^{2}$ Argelander Institut f\"{u}r Astronomie (AIfA),
  Auf dem H\"{u}gel 71, 53121 Bonn, Germany}

\pagerange{\pageref{firstpage}--\pageref{lastpage}} \pubyear{2011}

\maketitle

\begin{abstract}
  It is widely believed that star clusters form with low star
  formation efficiencies.  With the onset of stellar winds by massive
  stars or finally when the first super nova blows off, the residual
  gas is driven out of the embedded star cluster.  Due to this fact a
  large amount, if not all, of the stars become unbound and disperse
  in the gravitational potential of the galaxy.  In this context,
  \cite{Kroupa2002} suggested a new mechanism for the emergence of
  thickened Galactic discs. Massive star clusters add kinematically
  hot components to the galactic field populations, building up in
  this way, the Galactic thick disc as well.  In this work we perform,
  for the first time, numerical simulations to investigate this
  scenario for the formation of the galactic discs of the Milky
  Way. We find that a significant kinematically hot population of
  stars may be injected into the disk of a galaxy such that a thick
  disk emerges. For the MW the star clusters that formed the thick
  disk must have had masses of about $10^6\,M_\odot$.
\end{abstract}


\section{Introduction}
\label{sec:intro}

The formation of the Milky Way (MW) galaxy is a mystery unsolved yet.
Different models are trying to explain what were the initial
conditions that lead to the actual structure of the MW.  It is
commonly accepted that the structure of the Milky Way, and other
comparable disc galaxies, can be divided into three main components, the
bulge, the galactic spheroid and the disc.  The central bulge has a
mass of $\approx 10^{10}$~M$_{\odot}$ and a characteristic radius of
about $1$~kpc.  The galactic spheroid, which is also called the
stellar halo, has a mass of $\approx 3.7 \pm 1.2 \times
10^{8}$~M$_{\odot}$ \citep{bell08} and its mass is mostly confined
within the the Solar Radius.  The stellar halo contains globular
clusters, dwarf satellites and their tidal streams, that add up to a
mass of about $10^{6-7}$~M$_{\odot}$.

For the MW, the disc can be subdivided into at least two parts: the
thin disc with a mass of about $M_{\rm disc} = 5 \times
10^{10}$~M$_{\odot}$, that has exponential radial and vertical scale
lengths of approximately $h_{\rm R} = 2.3 \pm 0.6$~kpc \citep{hammer07}
and $h_{\rm z} \approx 300$~pc \citep{jur08} respectively.  The other
part is the thick disk which has
scale lengths of $h_{\rm thd,R} = 4.1 \pm 0.4$~kpc and $h_{\rm thd,z}
= 0.75 \pm 0.07$~kpc \citep{jong10}.  Near the Sun, the thick disc
comprises about $6$~per
cent of the thin disc mass, so that the thick disc mass amounts to
$M_{\rm thd} \approx 0.2-0.3 \times M_{\rm disc}$.  The thick disc is
made up mostly of low-metalicity  $([Fe/H] \leq -0.4)$ stars that
have a velocity dispersion perpendicular to the disc plane of
$\sigma_{\rm z,obs} \approx 40$~pc\,Myr$^{-1}$, compared to the
significantly smaller $\sigma_{\rm z}$ of the thin disc, which varies
from about $2-5$~pc\,Myr$^{-1}$ for the young stars \citep{Fuchs2001}.
These authors measured the velocity dispersion in the solar
neighbourhood as a function of age of the stars.  The oldest stars in
their sample (CNS4) have about $25$~pc\,Myr$^{-1}$ for $10$~Gyr old
stars.  But then this value might be 'contaminated' by thick disc
stars.

Several mechanisms have been proposed to explain the formation of the
thick disc in galaxies. One of these mechanisms was proposed by
\citet{aba03}, who suggest that the formation of the thick disc is the
direct accretion of stars from disrupted satellites.  The process of
accretion occurs approximately at coplanar orbits.  Another
explanation was suggested by \citet{ros08} and \citet{scho09}, who
consider the process of radial migration of the stars.  In this
mechanism, the stars which end up in the thick disc
are trapped onto a resonant co-rotation with spiral arms and may
migrate inwards and outwards along the spiral waves.  This process
conserves angular momentum and does not lead to significant heating of
the disc.  Another possible scenario is proposed by e.g.\
\citet{quin93}, \citet{kaz08} and \citet{villa08} and consists of the
thickening of a pre-existing thin disc through minor mergers.  The
thick disc is formed by the dynamical heating that is induced by
satellites merging with a primordial, rotationally supported thin
disc.  Finally, \citet{brok05} and \citet{bour07} suggest that the
formation of the thick disc is triggered in situ.  The process of star
formation occurs during/after gas rich mergers.

Each model explains different aspects and has its own implications for
the disc's kinematical and chemical properties. In the scenario of
\citet{aba03}, it is possible to distinguish two different dynamics
for the orbits of the stars, and they allow one to identify two
different components of the disc.  The thin disc is a kinematically
cold component with stars on circular orbits, and the thick disc is a
kinematically hot component with stars with orbital parameters
transitional between the thin disc and the spheroid.  The nature of
the population of stars in the thick disc comes from the tidal debris
of satellites, whose orbital plane is roughly coincident with the
disc.  Their orbits were circularized by dynamical friction prior to
their complete disruption. 

However, there is evidence for enhanced $\alpha$-elements abundances
in thick-disc stars.  This indicates a short star formation time-scale
in which enrichment is dominated by Type II supernovae (SNe II)
\citep{alves10,koba11}, rather than the slower time-scale expected for
dwarf galaxies.  In this case, one needs to consider the existence of
an active epoch of gas-rich mergers in the past history of galaxy,
with the bulk of the thick disc forming in situ \citep{brok05}, and
strong stellar scattering of clumps formed by gravitational
instabilities \citep{bour07} (rather than being accreted from
satellites) to explain these observations. 

During the last years understanding of star formation and star cluster
formation has progressed.  Now we know that almost all stars form in a
clustered mode \citep[e.g.][and follow-up publications]{lad03}.  The
star clusters form with low star formation efficiencies and,
therefore, they lose a large part of their stars that expand outwards
when residual gas is expelled by the action of massive stars.  In this
context, \citet{Kroupa2002} suggested another mechanism for the
formation of the thick disc.  Considering the star cluster formation
process described above, massive star clusters may add kinematically
hot components to galactic field populations, building up, in this
way, both galactic discs.

{  According to the \citet{Kroupa2002} model, the
  velocity dispersion of the stars in expansion (the new field
  population) is related with the mass of the initial star cluster,
  with its radius and also with the efficiency of star formation
  \citep[Eq.~1 in][]{Kroupa2002}.  It is proportional to the square
  root of the mass of the star cluster, such that heavier star
  clusters will generate a field population that presents higher
  dispersion velocities.  Thus, a simple estimate suggests that
  clusters of mass near $10^{5.5}$~M$_{\odot}$ may contribute new
  field populations that have a velocity dispersion of about
  $40$~pc\,Myr$^{-1}$ which is similar to the velocity dispersion of
  the thick disk.}   

In this work we perform numerical simulations to investigate this
scenario for the formation of both galactic discs of the Milky Way.
To accomplish this investigation we use star clusters of different
masses and we consider their disruption under distinct low star
formation efficiencies (we refer to such clusters as popping star
clusters).  We place the star cluster on an orbit around the Galactic
Centre at the Solar Radius of $8.5$~kpc and observe how the stars of
these star clusters distribute themselves in the potential of the MW.

In the next Section we describe the detailed setup of our
investigation and describe our results in Section~\ref{sec:res}.  We
discuss our findings in the last Section (\ref{sec:conc}).

\section{Simulations}
\label{sec:setup}

\subsection{Code Description}
\label{sec:code}

We simulate the popping star cluster using the particle mesh code
{\sc Superbox} \citep{Fellhauer2000}, which has moving high-resolution
sub-grids, which stay focused on the star cluster.  These sub-grids
provide high spatial resolution at the place of interest.  A
particle-mesh code neglects by default close encounters between
the particles (which are rather representations of the phase-space
than actual single stars) and is therefore called collision-less.
With a collision-less code it is possible to simulate galaxies without
having to use the actual number of stars ($10^{10}$). A code as
{\sc Superbox} is in principle unsuitable to simulate star clusters, as
with them two-body relaxation effects are important and govern their
evolution.  In our study we dissolve the star clusters immediately,
dispersing their stars within a short time (a few crossing times) into
the overall potential of the MW, therewith being in a regime where near
encounters between stars do not play any r\^{o}le.  Now the ability of
{\sc Superbox} to model particles with arbitrary masses gives us the
possibility to sample the phase space (e.g.\ by using more particles
than actual stars in the star cluster) more precisely.

The code is fast and resource-efficient, i.e.\ it requires a small
amount of computer memory.  Therefore, this code enables its user to
simulate objects with millions of particles, with high resolution, and
it can run on normal desktop computers.

{\sc Superbox} has two levels of high resolution sub-grids.  The
highest resolution grid has a resolution (i.e. cell-length) of
$0.2$~pc and covers the initially dense SC completely.  The medium
resolution grid has a cell-length of $40$~pc and covers the complete
$z$-height of interest.  Finally the outermost grid covers the
complete orbit of the SCs around the centre of the MW with a
resolution of $0.4$~kpc.

Regarding the fixed time-step of {\sc Superbox} we choose initially a
value to ensure that the SCs are kept stable without further
influence. Therefore the time-steps were different for the different
masses ranging from $0.03$~Myr for the $10^{4}$~M$_{\odot}$-clusters,
via $0.01$ and $0.006$~Myr to $100$~yr ($0.0001$~Myr) for the most
massive clusters ($10^{7}$~M$_{\odot}$).  As these tiny time-steps
require very long simulation times until we finally would reach
$10$~Gyr, we check the simulations and stop them as soon as the SCs
are completely dissolved.  When all stars are unbound and just
traveling under the influence of the global potential, we restart the
simulations with a larger time-step of $0.5$~Myr.

\subsection{Set-up}
\label{sec:sim-setup}

We consider the following set-up for our simulation. The potential of
the MW is modeled as an analytical background potential consisting of
a Hernquist sphere to generate the bulge,
\begin{equation}
  \Phi_{\rm bulge}(r)= - \frac{GM_{\rm b}}{r+a},
\end{equation}
using $M_{\rm b} = 3.4 \times 10^{10}$~M$_{\odot}$ and $a = 0.7$~kpc,
the Miamoto-Nagai model to mimic the disc,
\begin{equation}
  \Phi_{\rm disc}(r) = - \frac{GM_{\rm d}}
  {\sqrt{R^{2}+(b+\sqrt{z^{2}+c^{2}})^{2}}},
\end{equation}
 with $M_{\rm d} = 10^{11}M_{\odot}$, $b=6.5$~kpc and $c=0.26$~kpc.
 Finally, we use a logarithmic potential to account for the rotation
 curve of the MW disk,
\begin{equation}
  \Phi_{\rm halo}(r) = \frac{v^{2}_{0}}{2}\ln(r^{2} + d^{2})
\end{equation}
with $v_{0} = 186$~km\,s$^{-1}$ and $d = 12$~kpc.
The superposition of these components gives a good analytical
representation of the Milky Way potential today.  The potential of the
MW is kept constant throughout the simulation.  This is highly
idealised as we expect the Galaxy to grow and evolve over this period
of time.  We discuss this issue further in Sect.~\ref{sec:conc}.

\begin{table}
  \centering
  \caption{Initial conditions of the Plummer spheres which represent
    the star clusters for each simulations.  The first column shows
    the initial Plummer radius of the star clusters, the second the
    initial mass of the sphere and the third the initial
    characteristic velocity dispersion.  Finally the fourth and fifth
    columns represent the SFE and the final mass after gas expulsion.}
  \label{tab:init}
  \begin{tabular}{ccccc} \hline
    $R_{\rm pl}$ & $M_{\rm ecl}$ & $\sigma_{\rm ecl}$ & SFE & $M_{\rm f}$  \\
    $[$pc$]$ & $[$M$_{\odot}]$ & [$kms^{-1}$] & [\%] & $[$M$_{\odot}]$
     \\ \hline
     $ 1 $ & $ 2.2 \times 10^{4} $ & $ 5.28 $ & $ 40 $ & $ 8.80 \times
     10^{3} $   \\
    $ 1 $ & $ 2.2 \times 10^{4} $ & $ 5.28 $ & $ 20 $ & $ 4.40 \times
    10^{3} $   \\
    $ 1 $ & $ 2.2 \times 10^{4} $ & $ 5.28 $ & $ 1 $  & $ 0.22 \times
    10^{3} $   \\
    \hline
    $ 1 $ & $ 2.2 \times 10^{5} $ & $ 16.7 $ & $ 40 $ & $ 8.80 \times
    10^{4} $   \\
    $ 1 $ & $ 2.2 \times 10^{5} $ & $ 16.7 $ & $ 20 $ & $ 4.40 \times
    10^{4} $   \\
    $ 1 $ & $ 2.2 \times 10^{5} $ & $ 16.7 $ & $ 1 $  & $ 0.22 \times
    10^{4} $   \\
    \hline
    $ 1 $ & $ 2.2 \times 10^{6} $ & $ 52.8 $ & $ 40 $ & $ 8.80 \times
    10^{5} $   \\
    $ 1 $ & $ 2.2 \times 10^{6} $ & $ 52.8 $ & $ 20 $ & $ 4.40 \times
    10^{5} $   \\
    $ 1 $ & $ 2.2 \times 10^{6} $ & $ 52.8 $ & $ 1 $  & $ 0.22 \times
    10^{5} $   \\
    \hline
    $ 1 $ & $ 2.2 \times 10^{7} $ & $ 166.8 $ & $ 40 $ & $ 8.80 \times
    10^{6} $   \\
    $ 1 $ & $ 2.2 \times 10^{7} $ & $ 166.8 $ & $ 20 $ & $ 4.40 \times
    10^{6} $   \\
    $ 1 $ & $ 2.2 \times 10^{7} $ & $ 166.8 $ & $ 1 $  & $ 0.22 \times
    10^{6} $   \\
    \hline
    \end{tabular}
 \end{table}

The single cluster is represented by a \citet{Plummer1911} sphere.
Even though, this might not be the most ideal representation of a
young embedded star cluster, the Plummer distribution has the
advantage that all quantities are easily accessible with analytical
formulas.  And it is in fact a better representation of a young, newly
formed star cluster \citep{Kroupa2008}, which has not yet been tidally
truncated, than e.g.\ a King profile.  The Plummer profile has two
free parameters to choose, namely the Plummer radius, which we keep
fixed at $R_{\rm pl} = 1$~pc, and the total mass of the Plummer sphere
-- a parameter which is varied in this study.  All other quantities
(e.g.\ crossing-time, velocity dispersions, distribution of
velocities) are then simple functions of the two input parameters.
The choice of the small Plummer radius is based on observations of
embedded star forming regions \citep{Kroupa2008}.  Furthermore we
choose a cutoff radius of $5$~pc.  The choice of the cut-off radius is
justified by the fact that beyond $5 R_{\rm pl}$ there is only less
than $2$~per cent of the total mass of the Plummer model missing.
Each cluster has an initial mass of $M_{\rm ecl} = 2.2 \cdot 10^{4}$,
$2.2 \cdot 10^{5}$, $2.2 \cdot 10^{6}$ and $2.2 \cdot
10^{7}$~M$_{\odot}$ respectively leading to an initial crossing time
of $0.6$~Myr, $0.2$~Myr, $0.06$~Myr and $0.001$~Myr and is represented
with $2,500,000$ particles.

This configuration, placed on a circular orbit at the Solar Radius of
$8$~kpc, suffers from the process of gas expulsion.  This process is
mimicked by artificially reducing the mass of each particle according
to the SFE used.  We investigate a range of possible SFEs namely
$0.4$, $0.2$ and $0.01$.

Finally, we also use two distinct ways, how this mass-loss progresses.
We perform simulations with an instant mass-loss and simulations where
the mass in gas is lost during the time-interval of one crossing time
of the star cluster.

As the simulation time we choose a time-span of $10$~Gyr to ensure that
the particles have enough time to disperse into a stable configuration
in the Galactic potential.

\section{Results}
\label{sec:res}

\subsection{$z$-distribution of the stars}
\label{sec:space}

\begin{figure*}
  \centering
  \epsfxsize=8.5cm
  \epsfysize=6.5cm
  \epsffile{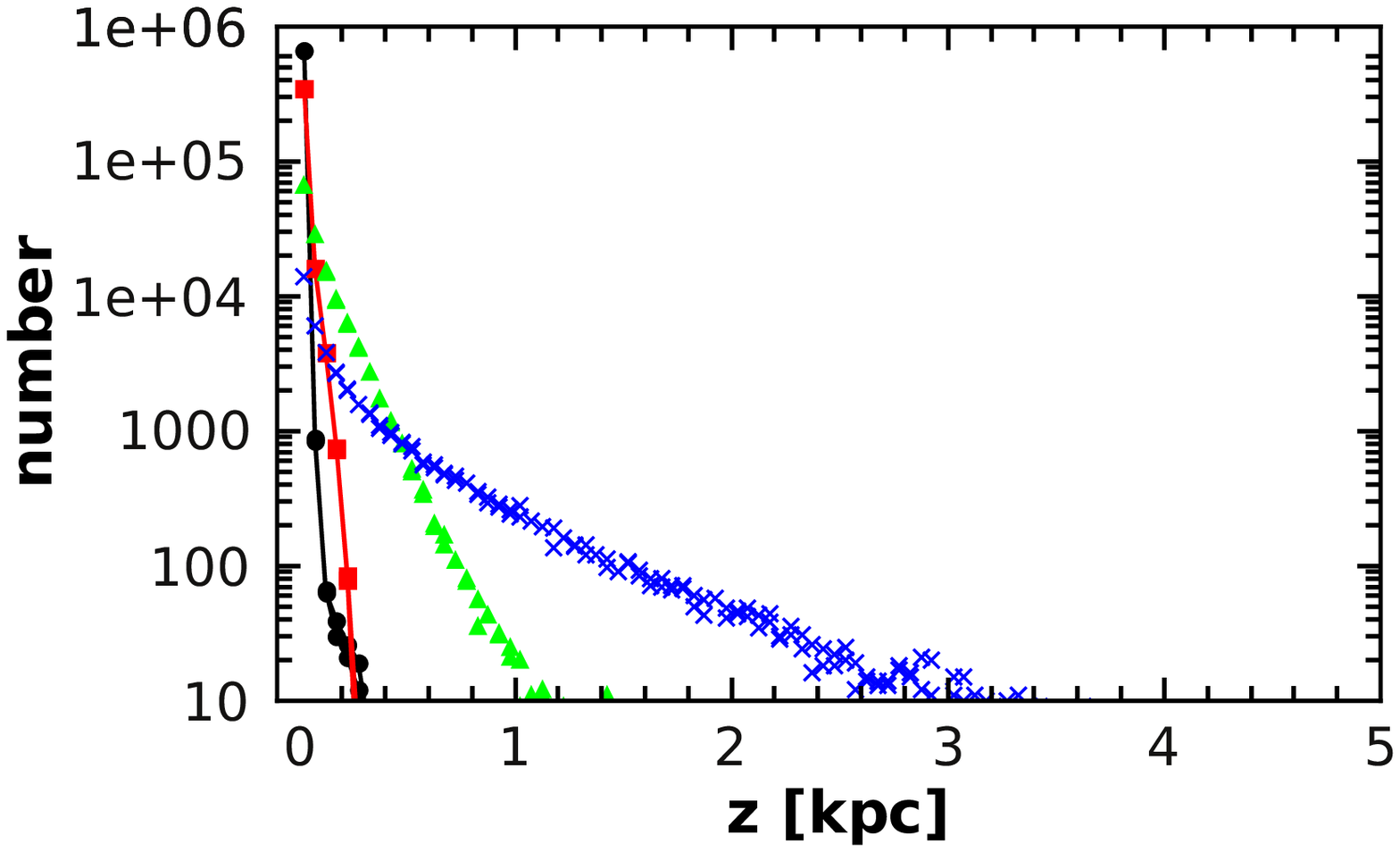}
  \epsfxsize=8.5cm
  \epsfysize=6.5cm
  \epsffile{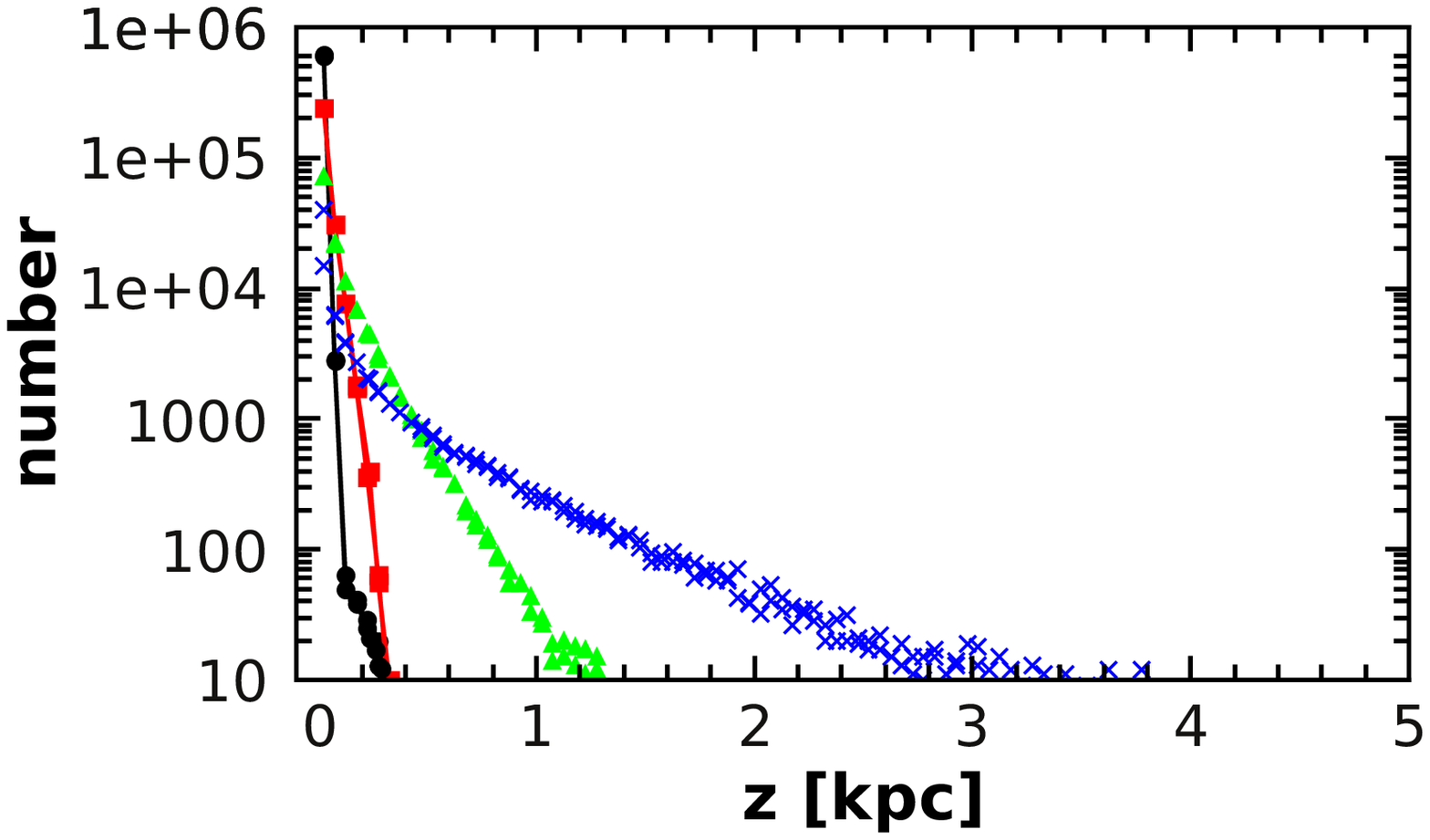}
  \epsfxsize=8.5cm
  \epsfysize=6.5cm
  \epsffile{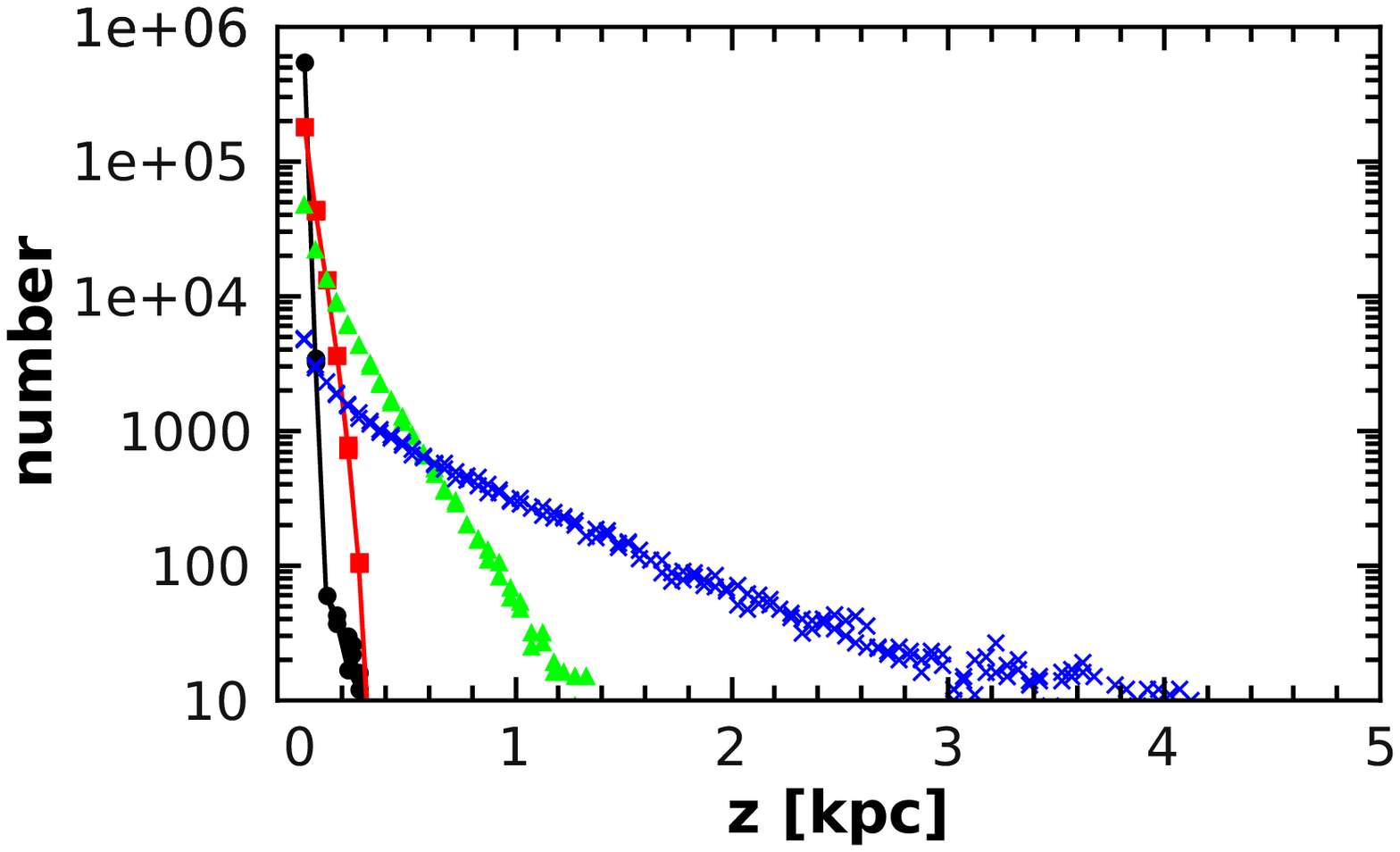}
  \epsfxsize=8.5cm
  \epsfysize=6.5cm
  \epsffile{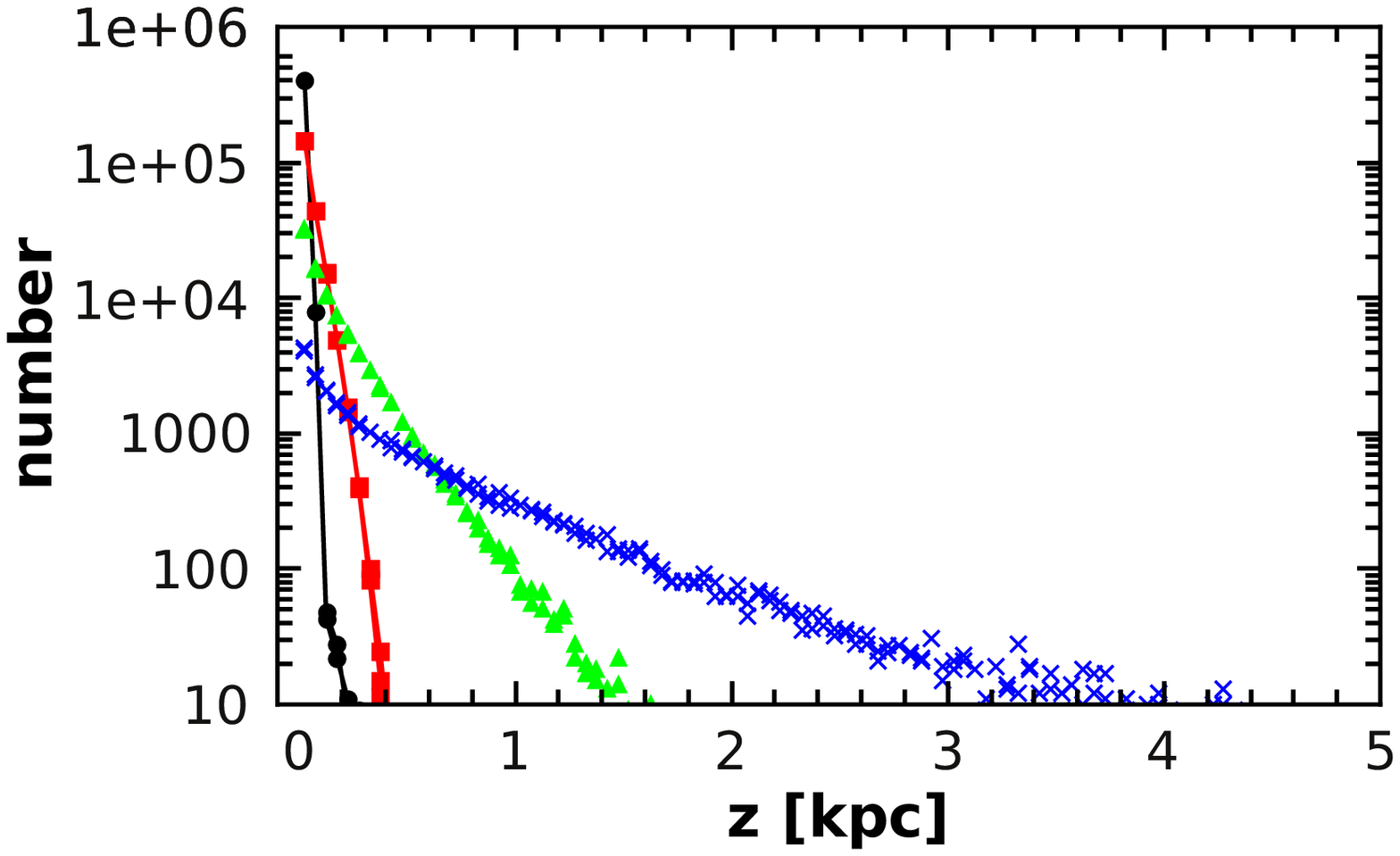}
  \epsfxsize=8.5cm
  \epsfysize=6.5cm
  \epsffile{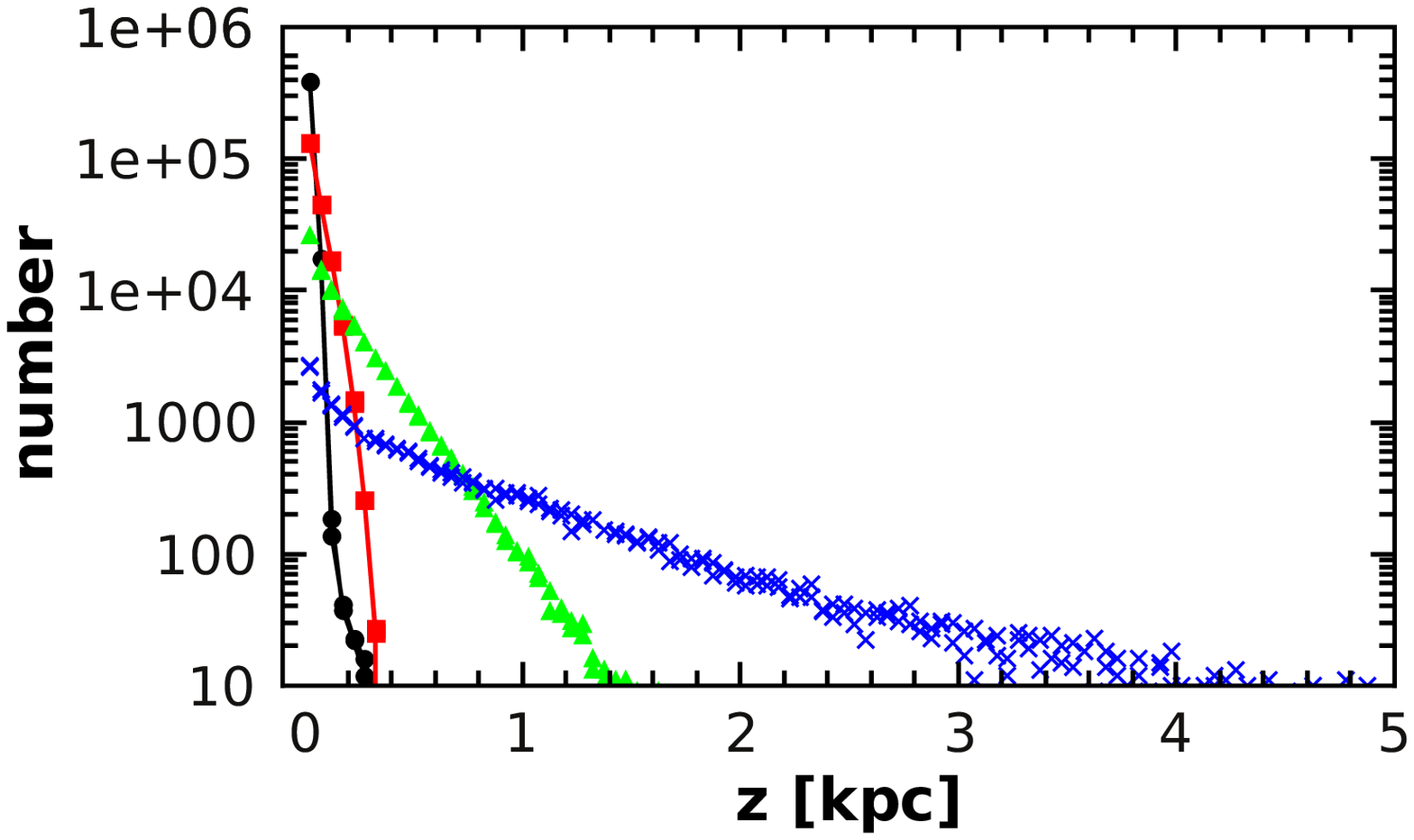}
  \epsfxsize=8.5cm
  \epsfysize=6.5cm
  \epsffile{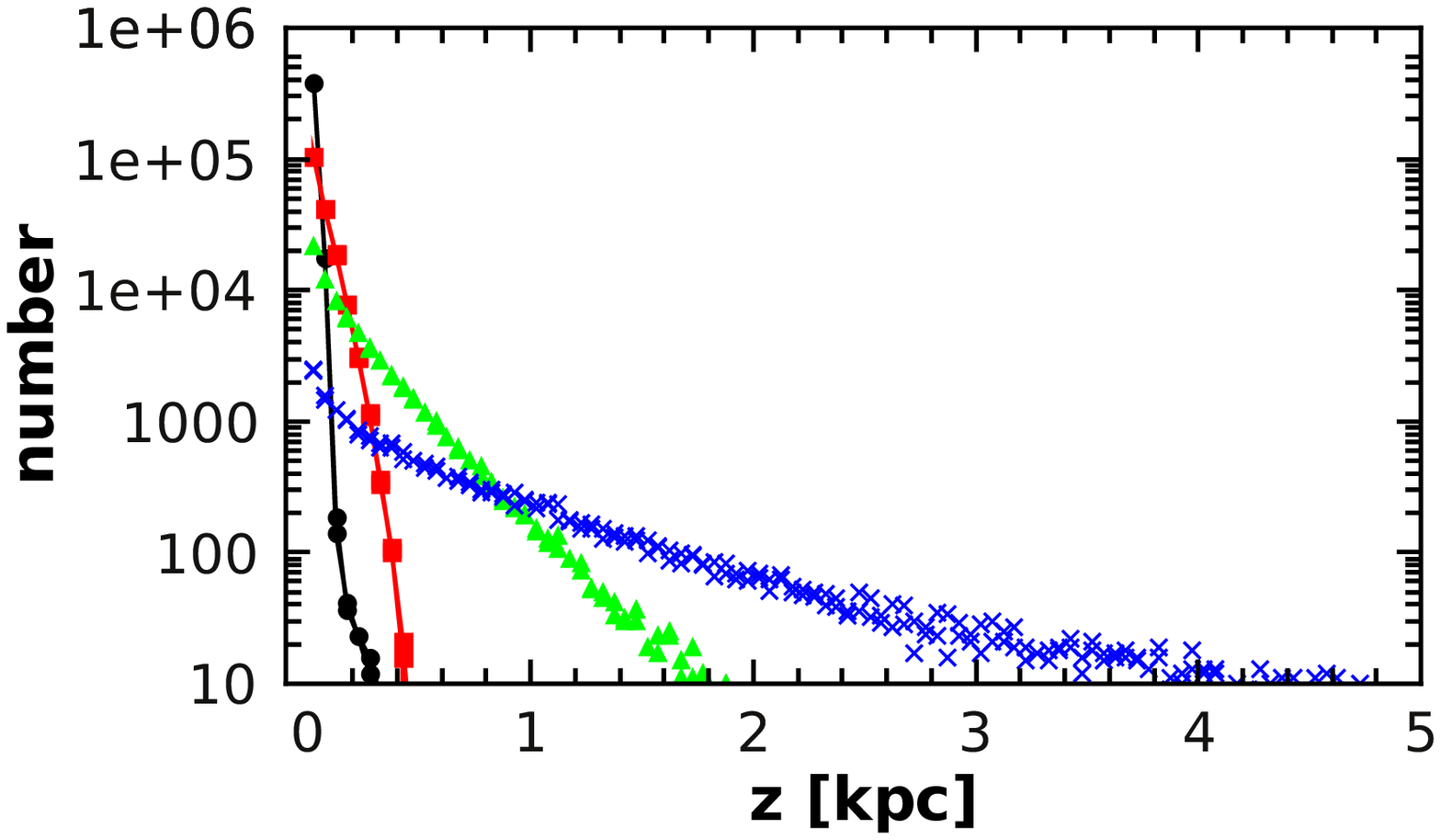}
  \caption{The $z$-distribution of particles of the dissolved star
    clusters after $10$~Gyr of simulation time.  From top to bottom we
    show the results of the different SFEs ($0.4$, $0.2$ and $0.01$).
    Left panels are for the gas-expulsion during a crossing-time and
    right panels show the results for instantaneous gas-expulsion.
    Inside each panel the curves from left to right represent the
    results of the different initial masses of the star clusters
    namely $M_{\rm ecl} = 2.2 \times 10^{4}$~M$_{\odot}$ (circles
    black on-line), $2.2 \times 10^{5}$~M$_{\odot}$ (squares red
    on-line), $2.2 \times 10^{6}$~M$_{\odot}$ (triangles green) and
    finally $2.2 \times 10^{7}$~$M_{\odot}$ (crosses blue).}
  \label{fig:hist-1}
\end{figure*}

\begin{table}
  \centering
  \caption{Results of the simulations.  The first three columns
    represent the initial mass of the star clusters, the SFE used and
    finally if the gas was expelled instantaneously $0$ or over a
    crossing-time $1$ (GET = gas expulsion time).  The fourth and
    fifth column represent the scale height of the thin and thick
    disc, respectively, after $10$~Gyr of simulation.  The last line
    represents our combined model (see main text).}
  \label{tab:fit}
  \begin{tabular}{cccccc} \hline
    $M_{\rm ecl}$ & SFE & GET & $h_{\rm z,thin}$ & $h_{\rm z,thick}$  \\
    $[$M$_{\odot}]$ & [\%] & [$t_{\rm cr}$] & [pc] & [pc]
     \\ \hline
    $2.2 \times 10^{4}$ & $40$ & $0$ & $9.0$ & ---   \\
    $2.2 \times 10^{4}$ & $40$ & $1$ & $10.1$ & ---   \\
    $2.2 \times 10^{4}$ & $20$ & $0$ & $13.0$ & ---   \\
    $2.2 \times 10^{4}$ & $20$ & $1$ & $10.0$ & ---   \\
    $2.2 \times 10^{4}$ & $1$ & $0$ & $16.4$ & ---   \\
    $2.2 \times 10^{4}$ & $1$ & $1$ & $16.0$ & ---  \\
    \hline
    $2.2 \times 10^{5}$ & $40$ & $0$ & $25.0$ & ---   \\
    $2.2 \times 10^{5}$ & $40$ & $1$ & $31$ & ---   \\
    $2.2 \times 10^{5}$ & $20$ & $0$ & $42.4$ & ---   \\
    $2.2 \times 10^{5}$ & $20$ & $1$ & $35.6$ & ---   \\
    $2.2 \times 10^{5}$ & $1$ & $0$ & $56.0$ & ---   \\
    $2.2 \times 10^{5}$ & $1$ & $1$ & $47.0$ & ---  \\
    \hline
    $2.2 \times 10^{6}$ & $40$ & $0$ & $27.0$ & $130$   \\
    $2.2 \times 10^{6}$ & $40$ & $1$ & $14.2$ & $110$   \\
    $2.2 \times 10^{6}$ & $20$ & $0$ & $35.2$ & $170$   \\
    $2.2 \times 10^{6}$ & $20$ & $1$ & $33.3$ & $149$   \\
    $2.2 \times 10^{6}$ & $1$ & $0$ & $39.0$ & $215$   \\
    $2.2 \times 10^{6}$ & $1$ & $1$ & $31.3$ & $183$  \\
   \hline
    $2.2 \times 10^{7}$ & $40$ & $0$ & $31.3$ & $500$   \\
    $2.2 \times 10^{7}$ & $40$ & $1$ & $51.3$ & $500$   \\
    $2.2 \times 10^{7}$ & $20$ & $0$ & $71.3$ & $568$   \\
    $2.2 \times 10^{7}$ & $20$ & $1$ & $70.0$ & $529$   \\
    $2.2 \times 10^{7}$ & $1$ & $0$ & $67.0$ & $676$   \\
    $2.2 \times 10^{7}$ & $1$ & $1$ & $76.0$ & $704$  \\
    \hline
    $8.8 \times 10^{7}$ & $20$ & $1$ & $17$ & $540$   \\
    \hline
    \end{tabular}
 \end{table}

In Fig.~\ref{fig:hist-1}, we show the number of stars with respect to
their $z$-height after $10$~Gyr of simulation time.  Therefore, we
count all stars in a vertical cylinder of $40$~pc radius around the
Sun.  To increase the statistical significance and also because we do
not know the actual position of the star cluster with respect to the Sun
(except that they have by design the same Galacto-centric distance),
we actually count the stars in an $80$~pc wide ($7.96\leq R =
\sqrt{x^{2}+y^{2}} \leq 8.04$~kpc) ring around the galaxy and plot the
$z$-height of the stars in Fig.~\ref{fig:hist-1}.  We choose to
simulate until $10$~Gyr to ensure that the objects had enough time to
evolve slowly into an equilibrium.  The SCs at this stage are
completely dissolved in almost all of the cases.

The different rows of Fig.~\ref{fig:hist-1} correspond to the three
different SFEs we adopt.  The top row shows the results of our SCs
having $40$~per cent of SFE, the middle row the results of the
$20$~per cent simulations and finally the bottom row the results
obtained with the extremely low efficiency of only $1$~per cent.  The
left column shows simulations where the gas-removal time was one
crossing-time of the star cluster, while the right column represents
the final distribution of stars when the gas was removed
instantaneously.  The curves, in each panel, from left to right
correspond to the star distributions of star clusters which had an
initial mass of $M_{\rm ecl} = 2.2 \times 10^{4}$~M$_{\odot}$ (black
on-line), $2.2 \times 10^{5}$~M$_{\odot}$ (red on-line), $2.2 \times 
10^{6}$~M$_{\odot}$ (green on-line) and finally $2.2 \times
10^{7}$~$M_{\odot}$ (blue on-line).

\begin{figure}
  \centering
  \epsfxsize=8.5cm
  \epsfysize=6.5cm
  \epsffile{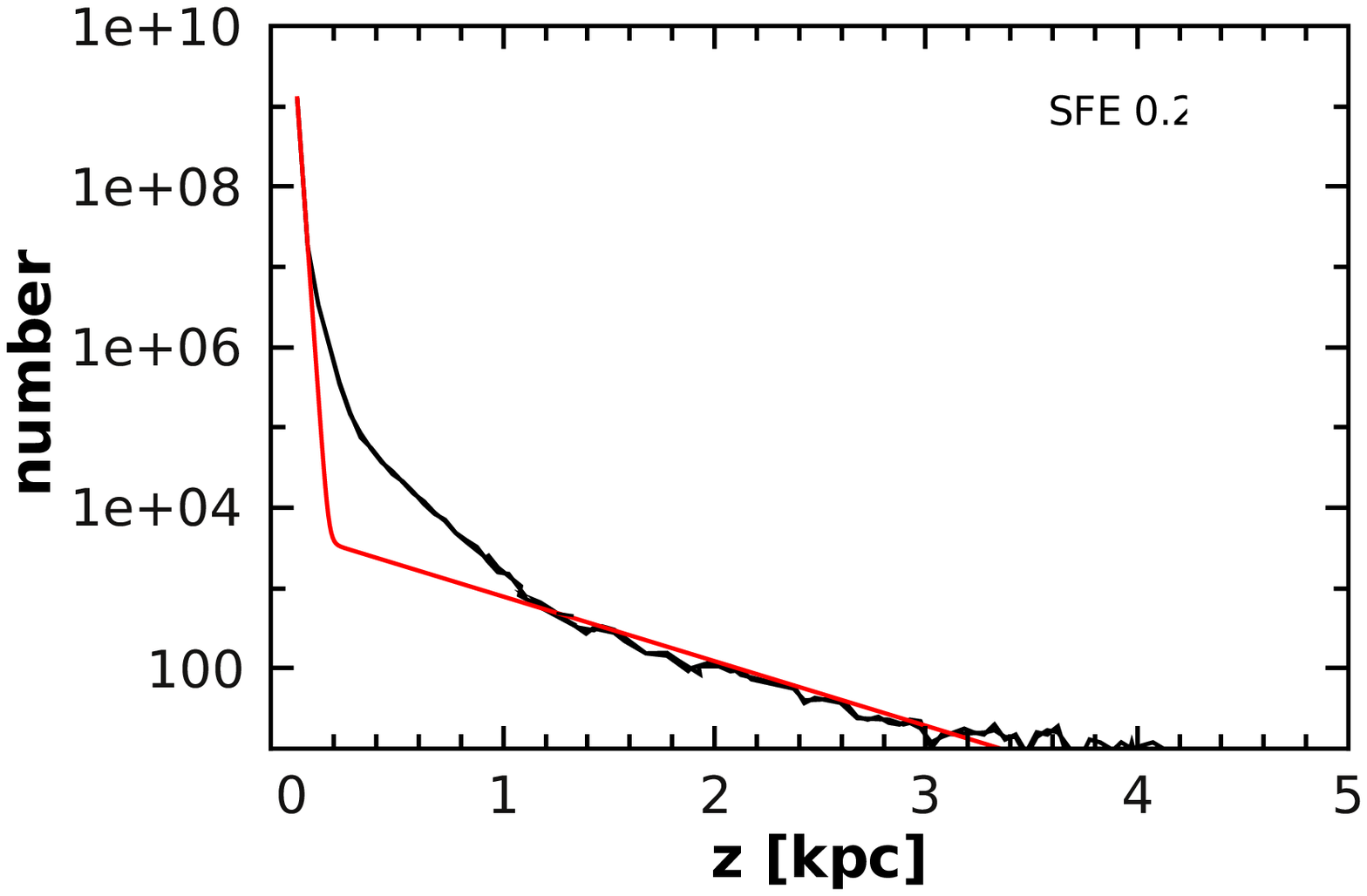}
  \caption{Distribution of stars in the $z$-direction if we add our
    different mass clusters according to a standard ICMF.  Fitting
    line is a double exponential.}
  \label{fig:hist-icmf}
\end{figure}

First we check our results for a dependency on the under-lying
analytic disc potential (Miamoto-Nagai model).  For this reason we
compare different fitting functions to our distributions.  One would
expect that if the under-lying potential had any influence it would be
visible in form of a similar distribution of the dispersed stars,
i.e.\ in our case a Plummer like distribution, as the Miamoto-Nagai
potential is a flattened version of the Plummer potential.  But our
results showed that the best fitting functions were either a single or
a double exponential.  Therefore we are quite confident, that our
results are not influenced by the particular choice of the analytical
potential.

The results show no difference with respect to the method of gas-
(mass-)loss we applied.  The final configurations are similar
irrespective if we remove the gas instantaneously or more smoothly
during a crossing-time of the star cluster.

But we see immediately a strong dependency on the initial mass of the
star cluster.  Low mass clusters ($2.2 \times 10^{4}$~M$_{\odot}$ and
$2.2 \times 10^{5}$~M$_{\odot}$) disperse their stars only close to
the Galactic plane and therefore can only contribute to the build-up
of the thin disc.  This changes as soon as we look at the high mass
clusters ($2.2 \times 10^{6}$~M$_{\odot}$ and $2.2 \times
10^{7}$~$M_{\odot}$).  These clusters are able to distribute their
stars out to $1$--$4$~kpc in $z$-height and therefore could be the
building blocks of the thick disc.

Finally the SFE has only a second order effect on the final
distribution.  The lower the SFE the wider is the spread of the final
particle distribution.  The reason is quite simple as a lower SFE
means that the initially binding potential of the embedded cluster is
deeper such that the stars, which amount to a mass $M$ say, move more
rapidly within it.  The same stellar mass, $M$, but confined by a
less-massive gas potential (being equivalent to a higher
star-formation efficiency), has a smaller velocity dispersion before
the gas is removed.  One also has to note that even though a SFE of
$0.4$ should in principal lead to a surviving remnant star cluster,
for all masses except our highest mass models, these remnants were not
massive enough to survive the whole $10$~Gyr of simulation time.  With
a SFE of 0.4 the cluster still 'pops' but  instead of dissolving
completely a small fraction of stars re-virialises into a bound star
cluster.  \citet{Kro2001} performed high-precision Nbody modelling of
this process explaining that a pre-Orion Nebula Cluster evolves into a 
Pleiades type cluster whereby $2/3$rds of the population "pop away".
The majority of stars thus gets distributed as unbound stars inside
the MW potential.  The fraction of stars which remains bound is a
function of the SFE.  For the low-mass cases this bound remnant does
not survive the tidal forces of the MW for very long.  Only the more
massive remnants stemming from massive initial clusters are able to
survive the full $10$~Gyr.  The future fate of these remnants is of no
concerns in our study, as our code is not able to describe the
internal dynamics of star clusters over long time intervals
correctly. 

As stated above we then fitted exponential functions to the obtained
star distributions in $z$-direction.  For all models we tried to fit a
single exponential of the form:
\begin{equation}
  \label{eq:single}
  \rho(z) = \rho_{\rm 0,thin} \cdot \exp \left(\frac{z}{h_{\rm z,thin}}
  \right),
\end{equation}
where $h_{\rm z,thin}$ represents the z-scale height of the thin disc
and compared the results with the ones obtained if we used a double
exponential function of the form:
\begin{equation}
  \label{eq:doble}
  \rho(z) = \rho_{\rm 0,thin} \cdot \exp \left(\frac{z}{h_{\rm z,thin}}
  \right) + \rho_{\rm 0,thick} \cdot \exp \left( \frac{z}{h_{\rm z,thick}}
    \right),
\end{equation}
where $h_{\rm z,thick}$ is the z-scale height of the thick disc.
The fitted values of $h_{\rm z,thin}$ and $h_{\rm z,thick}$ are shown
in Tab.~\ref{tab:fit}.  For the low mass clusters the best fit was
always the single exponential fit, while for the high mass clusters
the distribution was better fitted with a double exponential.  In the
table we only give the values for the best fit of the two fitting
functions adopted.  Again it is clear that the method of removing the
gas has no influence on the final distribution.  But we see a clear
trend towards larger scale-lengths by increasing the cluster mass and
by lowering the SFE.  For an example, in the case 
$M_{\rm ecl} = 2.2 \times 10^{6}$~M$_{\odot}$  and a SFEs of $40\%$ or
$20\%$, the z scale-height of the thin disc is about $14$~pc.  Using a
SFE of only $1$~per cent gives a z-scale height of $16$--$18.2$~pc.
The same type of behavior can be observed for the thick discs.

\begin{table}
  \centering
  \caption{$W$-velocity dispersions of our dissolving star clusters.
    The first three columns are the same as in Tab.~\ref{tab:fit}.  We
    then give the fitted values for the dispersion using the method
    described in the main text for the thin and thick disc component.
    The last column is the FWHM of the whole distribution.  Low mass
    star clusters, which only contribute to the thin disc have only one
    value fitted.  Last line is the result of our combined model
    using an ICMF.}
  \label{tab:vel}
  \begin{tabular}{ccccccc} \hline
    $M_{\rm ecl}$ & SFE & GET & $\sigma_{\rm z,thin}$ &$\sigma_{\rm
      z,thick}$ & FWHM  \\
    $[$M$_{\odot}]$ & [\%] & [$t_{\rm cr}$] & [km\,s$^{-1}$] &
    [km\,s$^{-1}$] & [km\,s$^{-1}$] \\
    \hline
    $2.2 \times 10^{4}$ & $40$ & $0$ & $1.04$ & --- & $1.0$ \\
    $2.2 \times 10^{4}$ & $40$ & $1$ & $0.78$ & --- & $1.0$ \\
    $2.2 \times 10^{4}$ & $20$ & $0$ & $1.59$ & --- & $1.2$ \\
    $2.2 \times 10^{4}$ & $20$ & $1$ & $1.24$ & --- & $1.0$ \\
    $2.2 \times 10^{4}$ &  $1$ & $0$ & $1.98$ & --- & $1.1$ \\
    $2.2 \times 10^{4}$ &  $1$ & $1$ & $1.98$ & --- & $1.1$ \\
    \hline
    $2.2 \times 10^{5}$ & $40$ & $0$ & $1.27$ & --- & $1.0$ \\
    $2.2 \times 10^{5}$ & $40$ & $1$ & $1.26$ & --- & $1.4$ \\
    $2.2 \times 10^{5}$ & $20$ & $0$ & $2.36$ & --- & $1.3$ \\
    $2.2 \times 10^{5}$ & $20$ & $1$ & $2.03$ & --- & $1.3$ \\
    $2.2 \times 10^{5}$ &  $1$ & $0$ & $2.80$ & --- & $1.3$ \\
    $2.2 \times 10^{5}$ &  $1$ & $1$ & $2.61$ & --- & $1.3$ \\
    \hline
    $2.2 \times 10^{6}$ & $40$ & $0$ & $3.28$ & $20.0$ & $5.9$ \\
    $2.2 \times 10^{6}$ & $40$ & $1$ & $4.10$ & $17.5$ & $2.3$ \\
    $2.2 \times 10^{6}$ & $20$ & $0$ & $4.50$ & $24.2$ & $1.5$ \\
    $2.2 \times 10^{6}$ & $20$ & $1$ & $4.01$ & $20.5$ & $1.5$ \\
    $2.2 \times 10^{6}$ &  $1$ & $0$ & $4.30$ & $27.3$ & $2.1$ \\
    $2.2 \times 10^{6}$ &  $1$ & $1$ & $4.55$ & $23.6$ & $1.7$ \\
   \hline
    $2.2 \times 10^{7}$ & $40$ & $0$ & $6.28$ & $58.8$ & $6.0$ \\
    $2.2 \times 10^{7}$ & $40$ & $1$ & $5.42$ & $59.2$ & $5.0$ \\
    $2.2 \times 10^{7}$ & $20$ & $0$ & $7.36$ & $58.3$ & $7.0$ \\
    $2.2 \times 10^{7}$ & $20$ & $1$ & $6.10$ & $50.2$ & $6.0$ \\
    $2.2 \times 10^{7}$ &  $1$ & $0$ & $12.9$ & $69.8$ & $12.0$ \\
    $2.2 \times 10^{7}$ &  $1$ & $1$ & $8.71$ & $69.3$ & $8.0$ \\
    \hline
    $8.8 \times 10^{7}$ & $20$ & $1$ & $1.26$ & $79.8$ & $1.26$ \\
    \hline
   \end{tabular}
 \end{table}

We should make clear that the actual distribution of stars is neither
a perfect single nor a perfect double exponential distribution.  With
the low-mass clusters we only see a thin distribution like a 'peak'
and only few stars in a sort of 'envelope' but with low $z$-heights.
Therefore a single exponential fits the data well.  For high-mass
clusters we see this 'peak' as well together with a very extended
structure (the 'envelope') reaching out to large $z$-heights.  These
clusters are able to spread out their stars to large radii.  For those
distributions a double exponential fits the data better.

Even though none of the values from Tab.~\ref{tab:fit} come anywhere
close to the values of the thin and thick disc of the MW, it is
astonishing that our high-mass clusters distribute their particles
automatically into a distribution which is best fitted by a double
exponential, i.e.\ show the same shape as the actual distribution of
MW stars.  The low numbers can easily be understood by the fact that
the MW is not made out of one single dispersed star cluster at one
certain time and on one certain orbit.  We rather have an overlay of
star clusters forming with different masses at different times and on
all radial distances.  Furthermore, we do not take any further effect
into account which could enhance the vertical spread of the star
particles, e.g. scattering with giant molecular clouds or spiral
arms, adiabatic heating due to the growth in mass of the MW disk.

\subsubsection{Combined model}
\label{sec:comb-z}

To at least asses the influence a generation of star clusters would
have onto the results we build our combined model by assuming an
initial cluster mass function (ICMF).  The ICMF is a fundamental
property of the process of star formation in galaxies \citep{kroup02}.
It gives the number of star clusters in a certain mass interval
$dM_{\rm ecl}$ which form during one event of star formation,
\begin{equation}
  N(M_{\rm ecl}) \propto M_{\rm ecl}^{-\beta}dM_{\rm ecl},
\end{equation}
where $\beta$ is the spectral index of cloud mass spectrum.  In this
work we use $\beta=2$, because this value was measured for young star
clusters in the Antennae \citep{whit99} and generally by \citet{weid04}.
Because we use a particle-mesh code, where particles represent
phase-space elements and not single stars, all of our model clusters
have the same number of particles, irrespective of their mass.  We
therefore are able mimic an ICMF with a power law index of $\beta=2$
by co-adding our results using a power of~$\beta-1=1$ only, to ensure
that every particle has the correct weight in the combined model.

The density distribution of the disrupted star clusters at $10~Gyr$
using the ICMF is shown in Fig~\ref{fig:hist-icmf}.  In this figure we
consider star clusters with a SFE of $20\%$ and where the gas
expulsion happens during a crossing time.  One clearly sees that the
shape of the density distribution has a similar resemblance as the
thin and thick disc of the MW.  We fit a double exponential function
given by Eq.~\ref{eq:doble} and obtain values for the $z$
scale-heights of $h_{\rm z,thin} = 50$~pc for the thin disc and
$h_{\rm z,thick} = 500$ for the thick disc (see also last line of
Tab.~\ref{tab:fit}).  These values are now much closer to the real
values of the MW and therefore we believe that by adding also
additional scattering and multiple star formation events, we may be
able to account for the real discs of the MW.

\subsection{$W$-velocity distributions}
\label{sec:vel}

\begin{figure}
  \centering
  \epsfxsize=8.5cm
  \epsfysize=6.5cm
  \epsffile{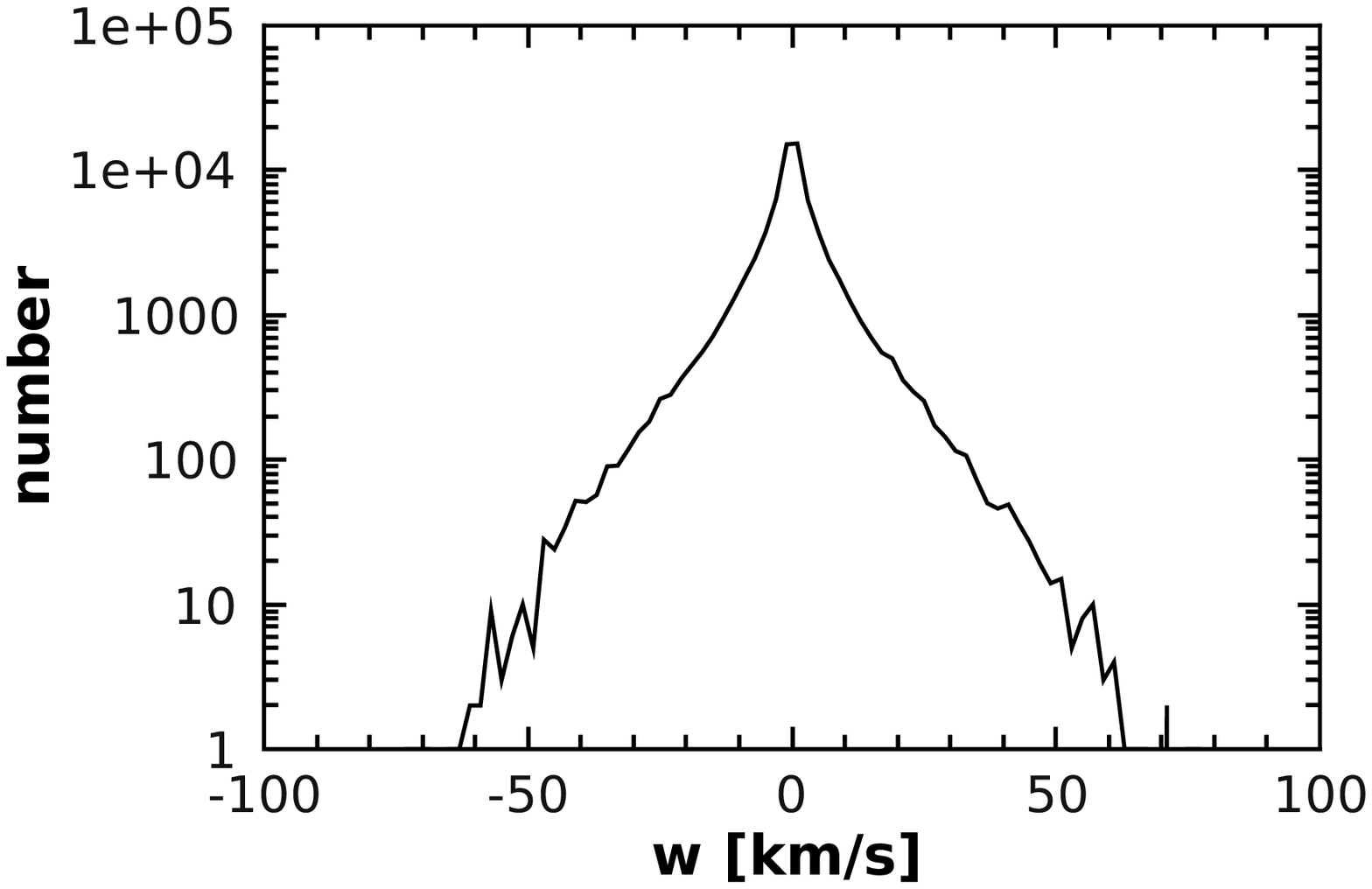}
  \epsfxsize=8.5cm
  \epsfysize=6.5cm
  \epsffile{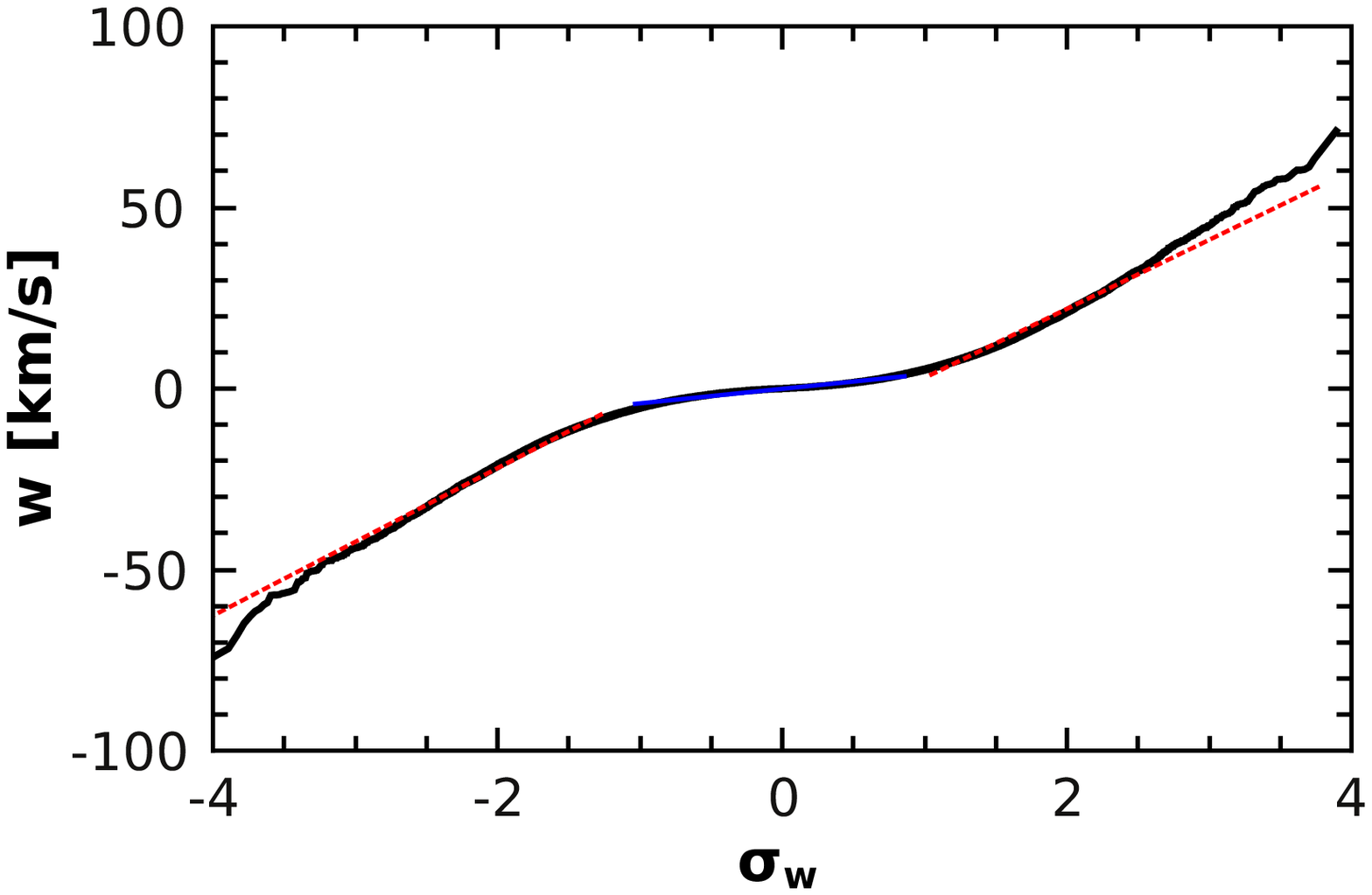}
  \caption{In the top panel we show the $W$-velocity distribution of
    one of our high mass models (see main text).  One clearly sees the
    non-Gaussianity of the distribution.  The lower panel shows the
    probability plot for this distribution with two straight lines
    fitted to the inner and outer parts respectively, giving the
    velocity dispersion for the two parts.}
  \label{fig:vel-m06}
\end{figure}

\begin{figure}
  \centering
  \epsfxsize=8.5cm
  \epsfysize=6.5cm
  \epsffile{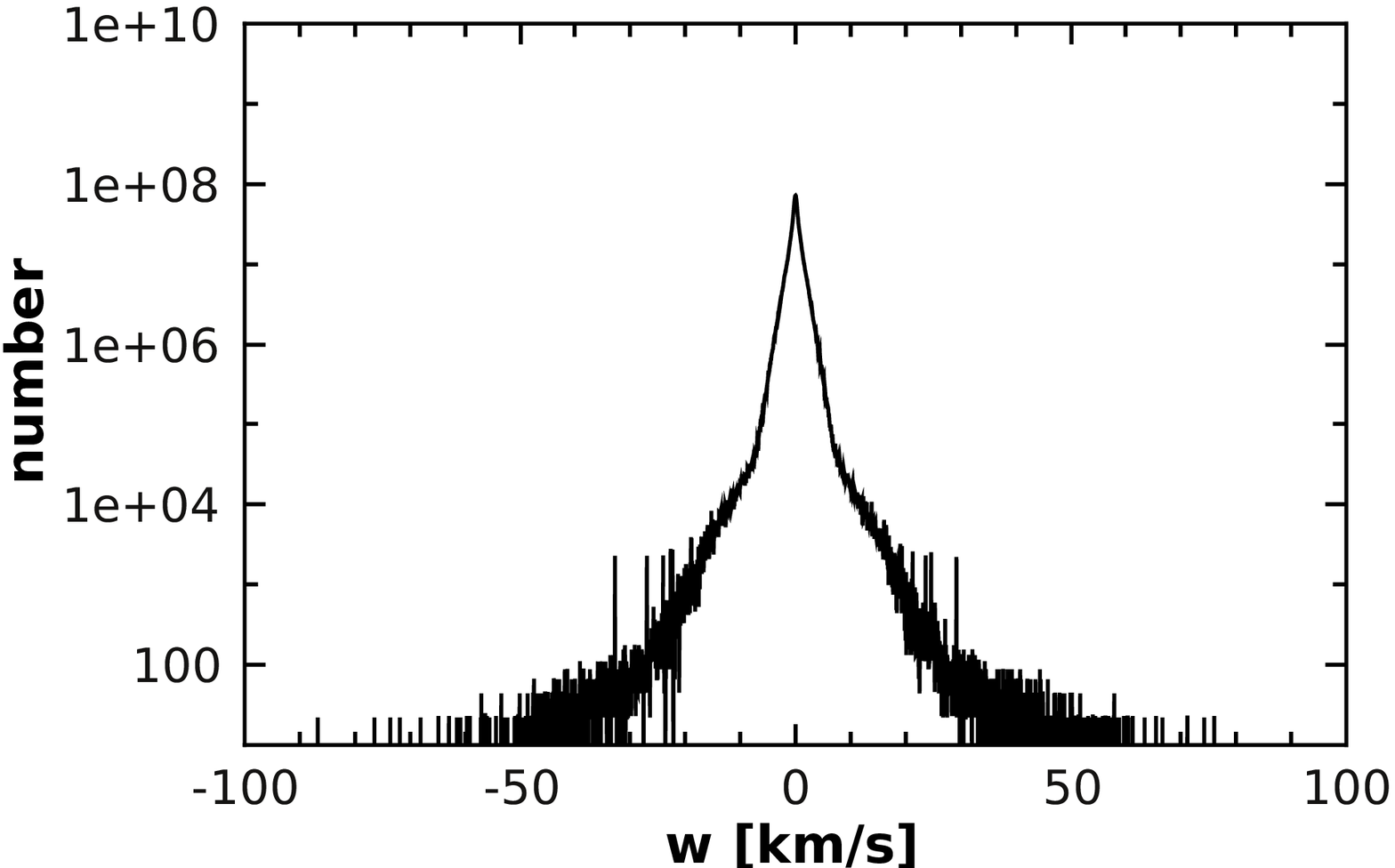}
  \epsfxsize=10.0cm
  \epsfysize=7.0cm
  \epsffile{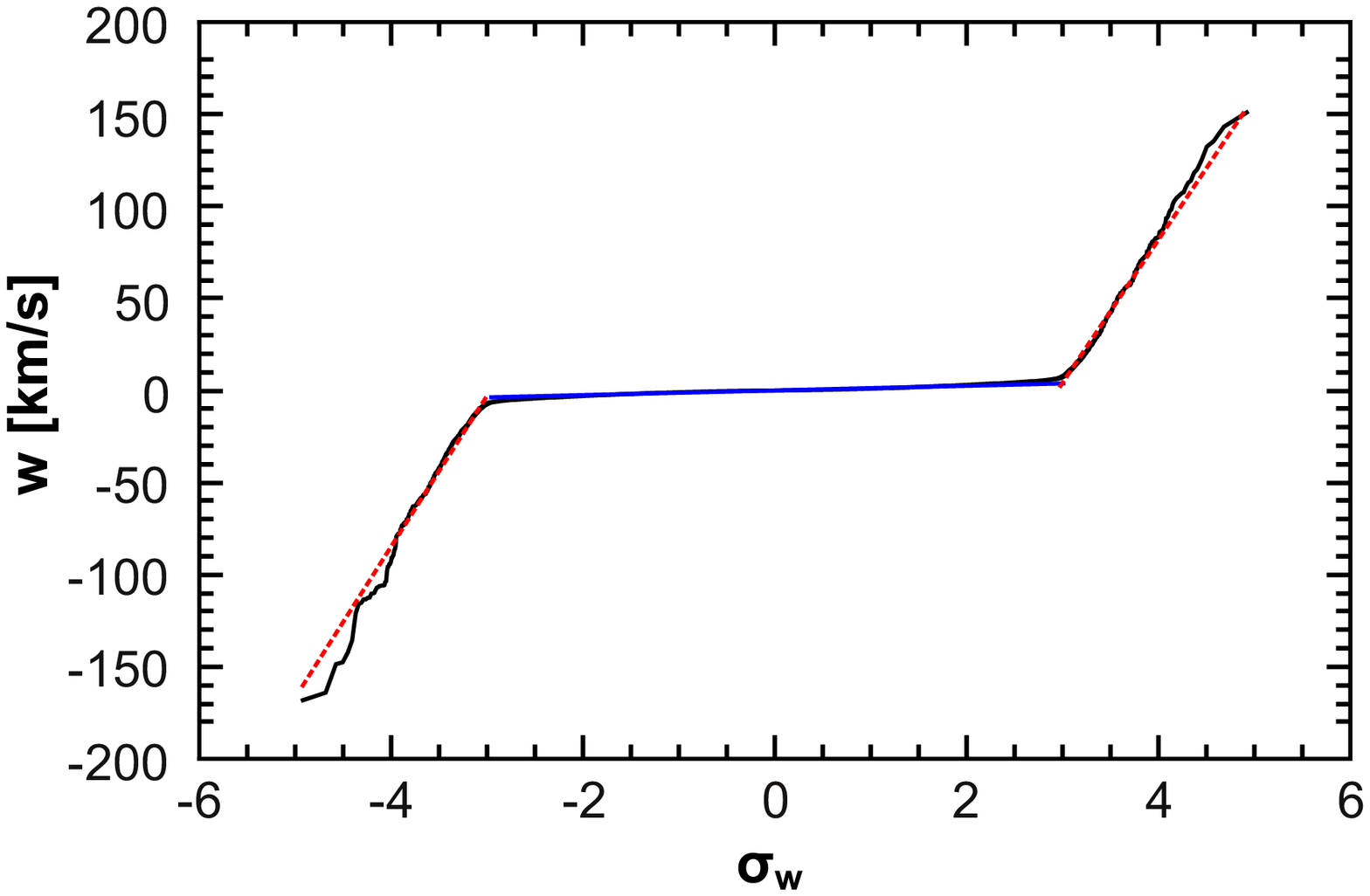}
  \caption{Same as Fig.~\ref{fig:vel-m06} but now for our combined
    model using an ICMF as described in the main text.}
  \label{fig:vel-icmf}
\end{figure}

In addition to studying the shape of the $z$-distribution of the
stars, we can also use our simulations to look at the velocity space.
Table~\ref{tab:vel} shows the velocity dispersion obtained for all our
simulations by fitting straight lines to the probability plot of
$W$-velocities (velocity in $z$-direction), as described in
\citet{boch07}.  The probability plot is a graphical technique where
we plot the cumulative probability distribution in units of the
standard deviation of the distribution.  When the distribution is a
Gaussian, it shows as a straight line in the probability plot with a
slope corresponding to the standard deviation.  The $y$-axis-intercept
is equal to the median of the distribution.  When the distribution is
non-Gaussian, the plotting line deviates from a straight line.  We are
therefore able to fit two separate lines to the inner and outer parts
of the distribution, resembling the velocity dispersion for the thin
and thick disc component of our models.  Again, we note that our
low-mass models only contribute to the thin disc component.

As example we show the velocity distribution and its probability plot
for a high mass model in Fig.~\ref{fig:vel-m06}.  It shows a star
cluster with a SFE of $20$~\% and a mass of $2.2 \times
10^{6}$~M$_{\odot}$.  Clearly, the velocity distribution of the stars
is not a Gaussian at all.  Also two Gaussians will not represent this
kind of distribution fully.  But it shows clear similarities with the
velocity distribution of stars found in the MW \citep{boch07}.  So
even though we have an overlay of an undefined number of Gaussians we
perform the same analysis as observers and fit straight lines to the
inner and outer parts of the probability plot.  The probability plot
is well fitted with two lines: low dispersion velocities ($< \mid1
\sigma \mid$), from the population of stars confined to the thin disc,
and high dispersion velocities ($> \mid1 \sigma \mid$), where the
population of the stars corresponds to the thick disc. In our example,
the slope for the velocity dispersion for the stars of the thin disc
is $4.01$~km\,s$^{-1}$ and for the thick disc stars
$20.5$~km\,s$^{-1}$. 

Additionally we measure the full width half maximum of the velocity
distribution and give these values as the last entry in
Tab.~\ref{tab:vel}.

{  If we compare the velocity dispersions of the initial
  star clusters with the dispersions we measure of the dispersed stars
  we see the following effects:
  \begin{itemize}
  \item The dispersion of the dispersed stars is always significantly
    lower than the initial dispersion inside the star cluster.  
  \item The ratio of final to initial velocity dispersion decreases
    with increasing mass (i.e.\ initial dispersion) of the star
    cluster.
  \item This fraction is higher for the stars in the thick disc
    component. 
  \item We get higher dispersions for a given initial cluster mass, if
    the SFE is lower.  The reason is that the stars at lower SFEs have
    a smaller potential barrier to cross.
  \item If the initial mass of the star cluster is about
    $10^{6}$~M$_{\odot}$ the velocity distributions are better fitted
    by two Gaussians than just one; similar to the $z$-distributions. 
  \end{itemize}
}

\begin{table}
  \centering
  \caption{  Ratio between initial velocity dispersion of
    the star clusters and the final velocity disspersion of the
    dispersed stars.  Shown are the models with a SFE of $0.01$ and 
    instanteneous gas-expulsion only.}
  \label{tab:dispcomp}
  \begin{tabular}{crrc}
    initial mass & final $\sigma$ & initial $\sigma$ & ratio \\
    $[$M$_{\odot}$] & [km\,s$^{-1}$] & [km\,s$^{-1}$] & \\ \hline
    $2.2 \times 10^{4}$ &  $2.0$ &   $5.3$ & $0.38$ \\
    $2.2 \times 10^{5}$ &  $2.8$ &  $16.7$ & $0.17$ \\
    $2.2 \times 10^{6}$ & $27.3$ &  $52.8$ & $0.52$ \\
    $2.2 \times 10^{7}$ & $69.8$ & $166.8$ & $0.42$ \\
    \hline
  \end{tabular}
\end{table}

\subsubsection{Combined model}
\label{sec:comb-w}

As with the $z-$distribution we now investigate how our results change
if we adopt an ICMF.  We combine our models the same way as done for
the $z$-distribution and show the combined velocity distribution in
the top panel of Fig.~\ref{fig:vel-icmf} (and also in the last line of
Tab.~\ref{tab:vel}).  The derived probability plot to this
distribution is shown in the lower panel of Fig.~\ref{fig:vel-icmf}.
Fitting straight lines to the inner and outer parts give velocity
dispersions of $1.3$ and $80$~km\,s$^{-1}$ for the thin and thick disc
component respectively.

As we can see, the velocity dispersion is much larger in the thick
disc than in the thin disc.  Comparing our values of the solar
neighborhood with published observations
\citep{bens03,vall06,veltz08}, which report $\sigma_{\rm w,thick} =
38$~km\,s$^{-1}$ and $\sigma_{\rm w, thin} = 16$~km\,s$^{-1}$ for the
thick disc and thin disc respectively, shows that we underestimate the
thin disc values quite substantially and overestimate the value for
the thick disc.  An explanation could be that we do not take any other
scattering mechanisms into account, which would elevate our thin disc
dispersion.  For the thick disc it could probably be, that there never
were any SCs with $10^{7}$~M$_{\odot}$ formed in the history of the
MW.  These high-mass clusters are mainly responsible for the elevated
velocity dispersion in the thick disc component of our models.

\subsection{Of Christmas trees and azimuthal fish}
\label{sec:various}

\begin{figure}
  \centering
  \epsfxsize=8.5cm
  \epsfysize=6.5cm
  \epsffile{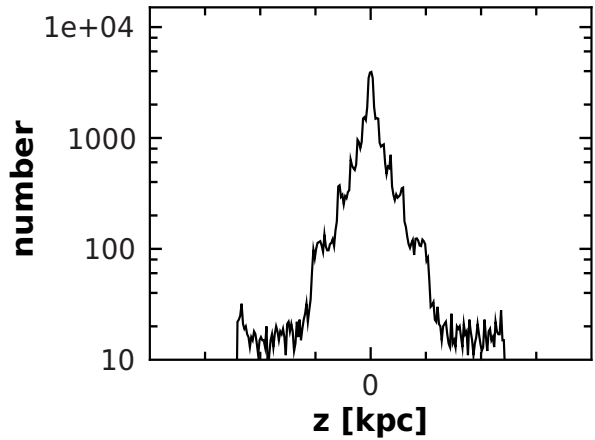}
    \caption{Example for a Christmas tree as found in all of our
      simulations.  It is a peculiar transient feature seen only after
      one revolution around the galaxy $\approx 270$~Myr (also the
      simulation time shown in the figure).  It disappears after a few
      revolutions more.}
  \label{fig:chris2}
\end{figure}

One interesting result can be observed in our simulation when we focus
on the $z$-distribution of the stars at early stages ($< 1$~Gyr) of
the SC evolution.  For an example, in Fig~\ref{fig:chris2} we show the
$z$-distribution of the stars, after $270$~Myr, for the simulation
with an initial mass of $2.2 \times 10^{7}$~M$_{\odot}$ and a SFE of
$0.2$.  

\begin{figure*}
  \centering
  \epsfxsize=8.5cm
  \epsfysize=6.5cm
  \epsffile{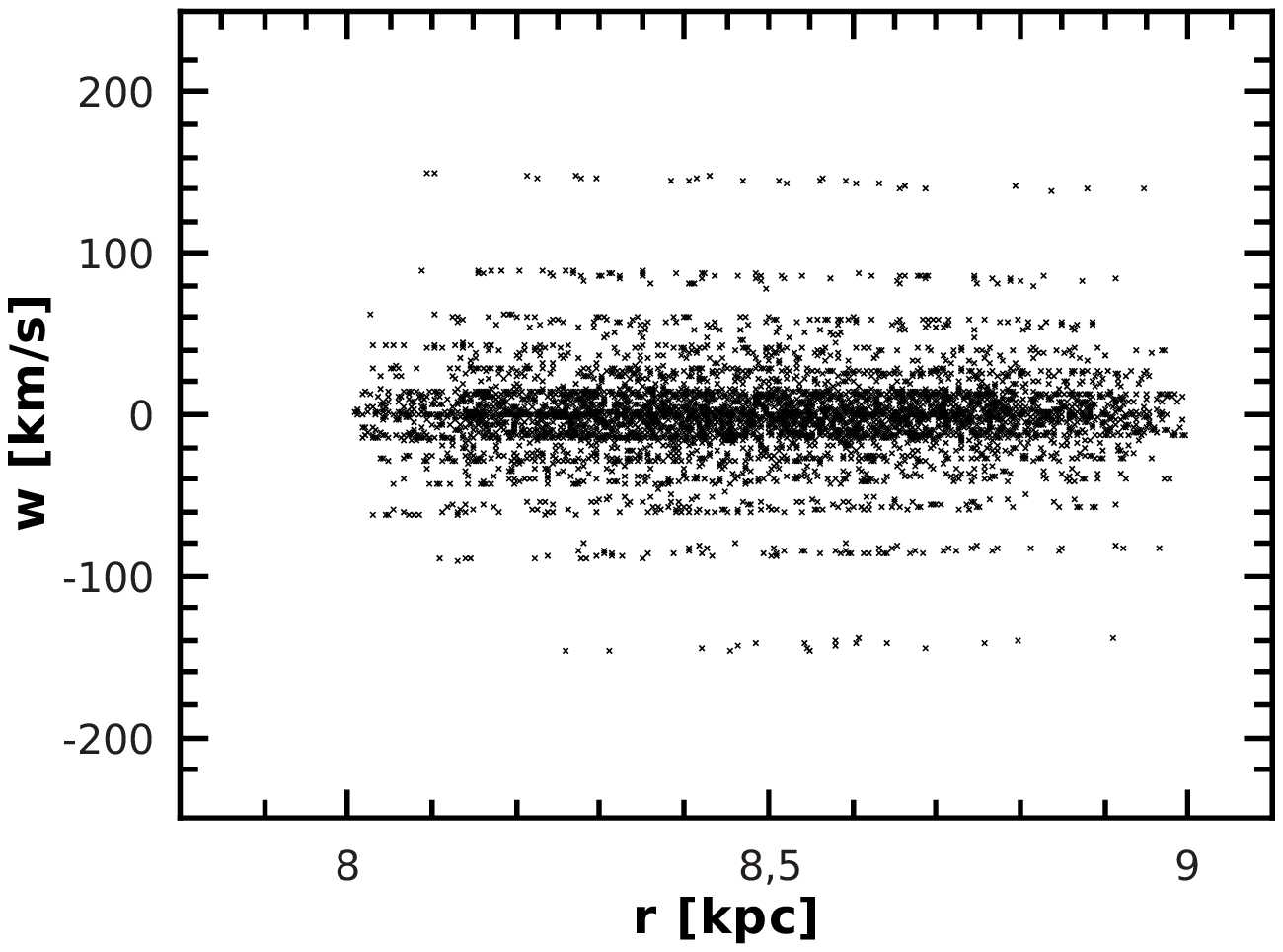}
  \epsfxsize=8.5cm
  \epsfysize=6.5cm
  \epsffile{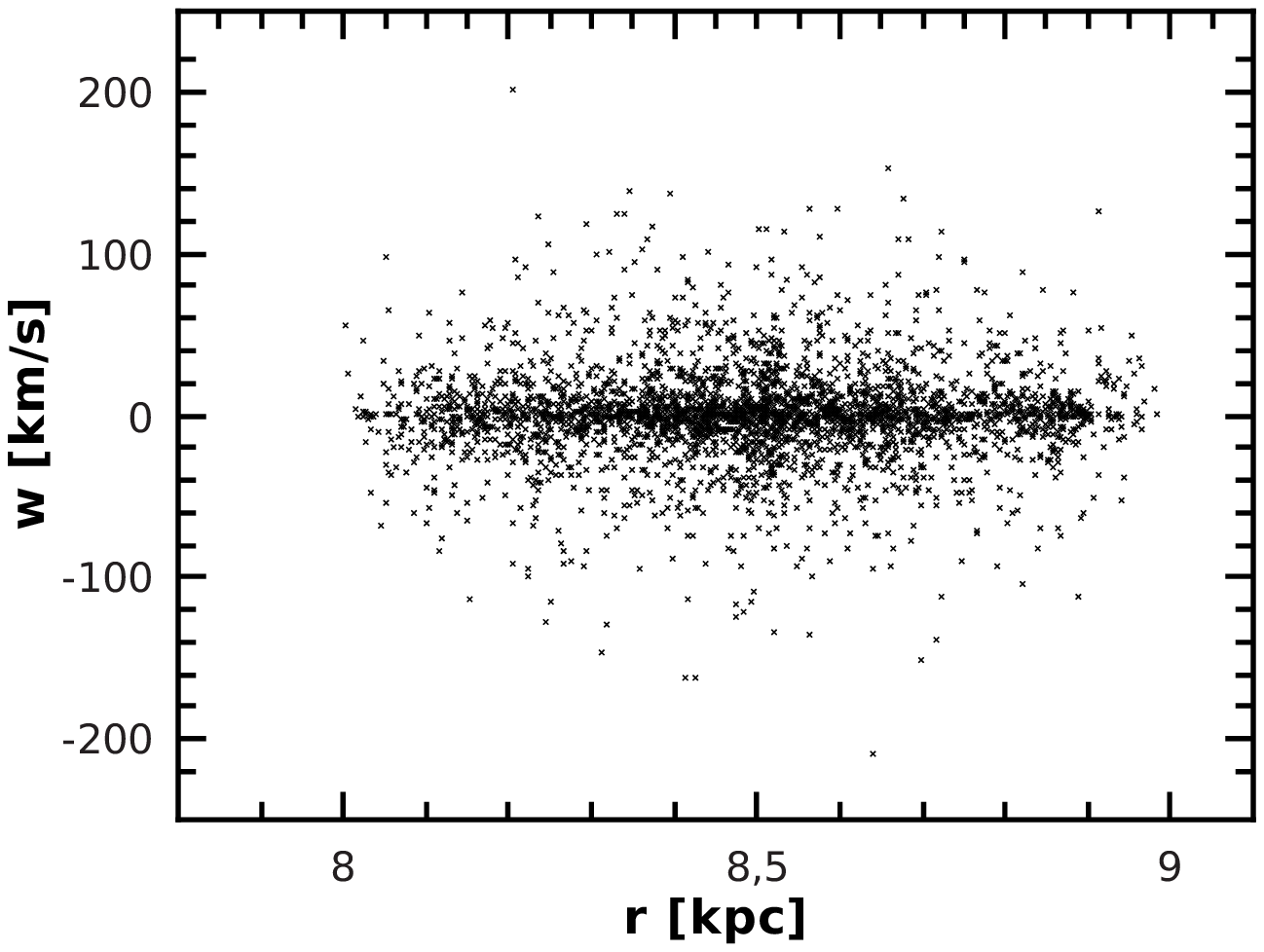}
  \epsfxsize=8.5cm
  \epsfysize=6.5cm
  \epsffile{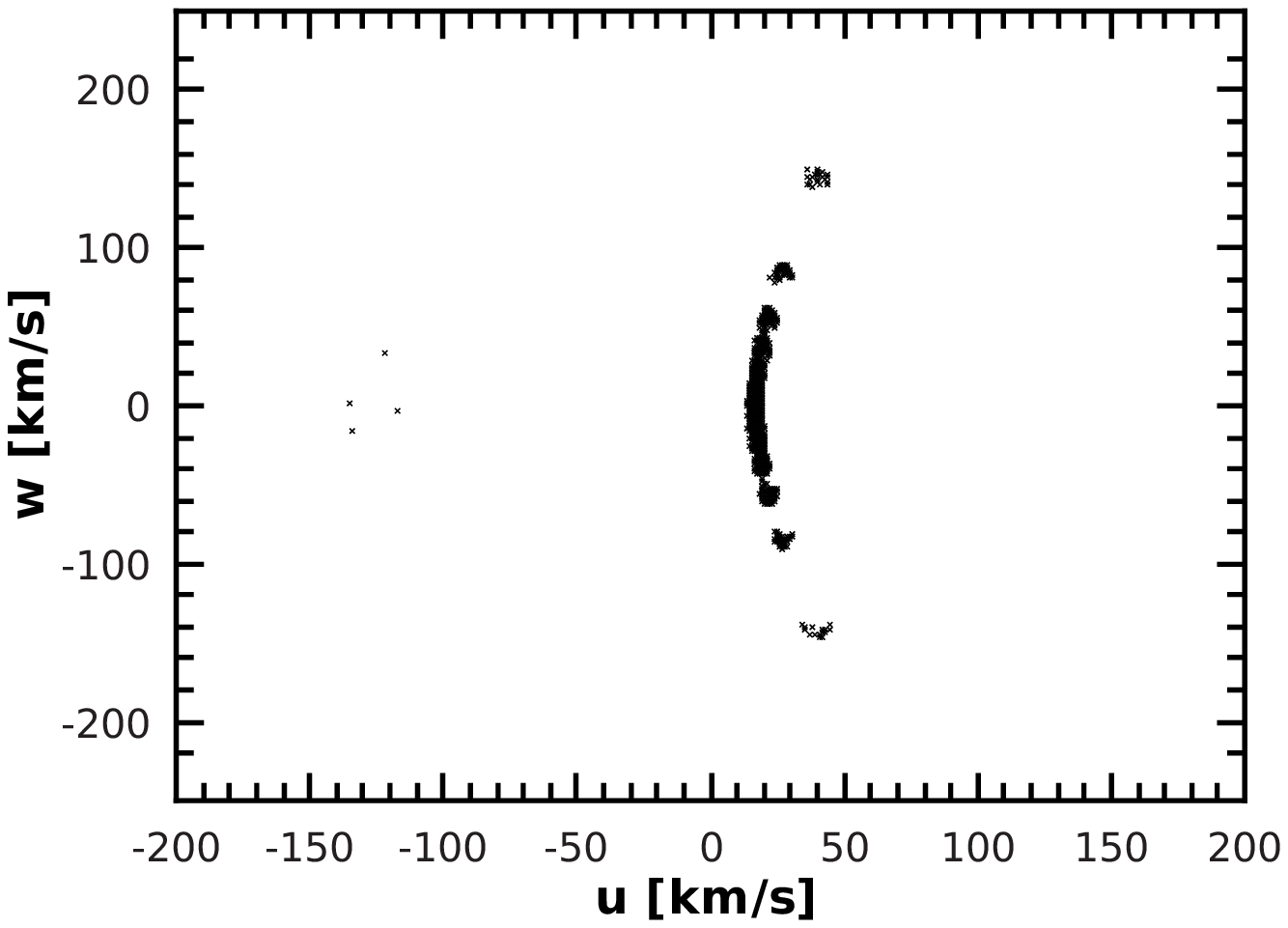}
  \epsfxsize=8.5cm
  \epsfysize=6.5cm
  \epsffile{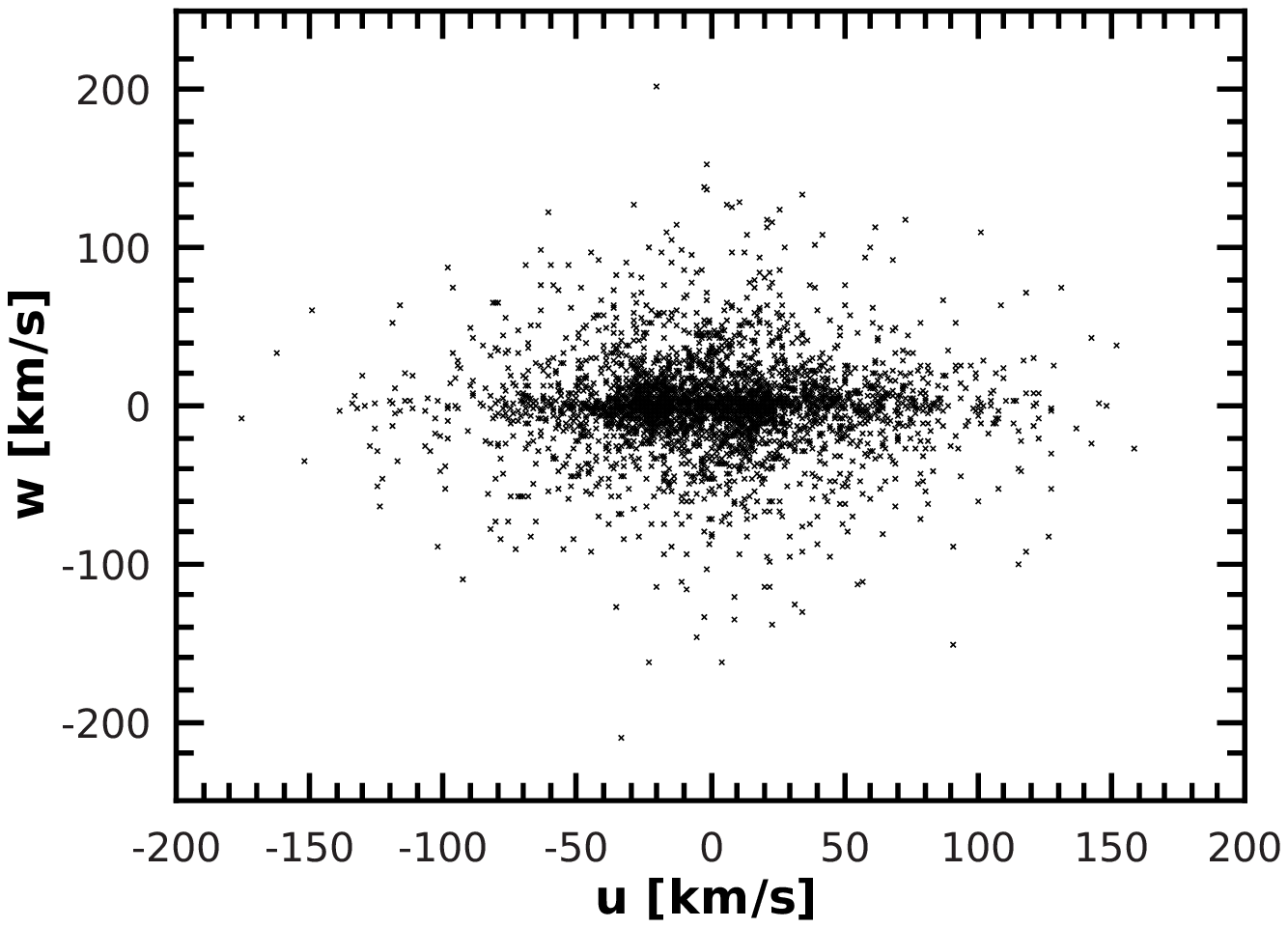}
  \epsfxsize=8.5cm
  \epsfysize=6.5cm
  \epsffile{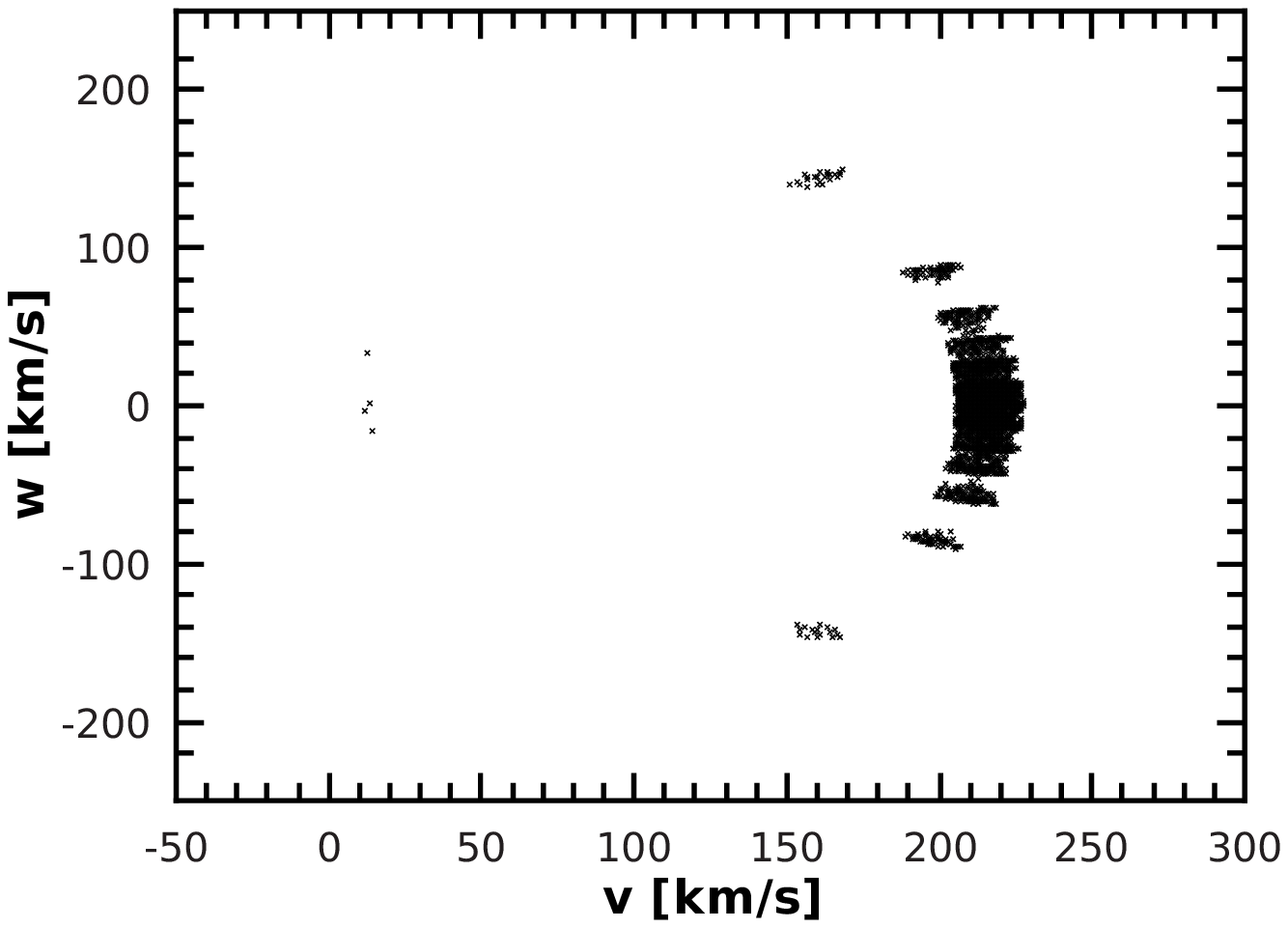}
  \epsfxsize=8.5cm
  \epsfysize=6.5cm
  \epsffile{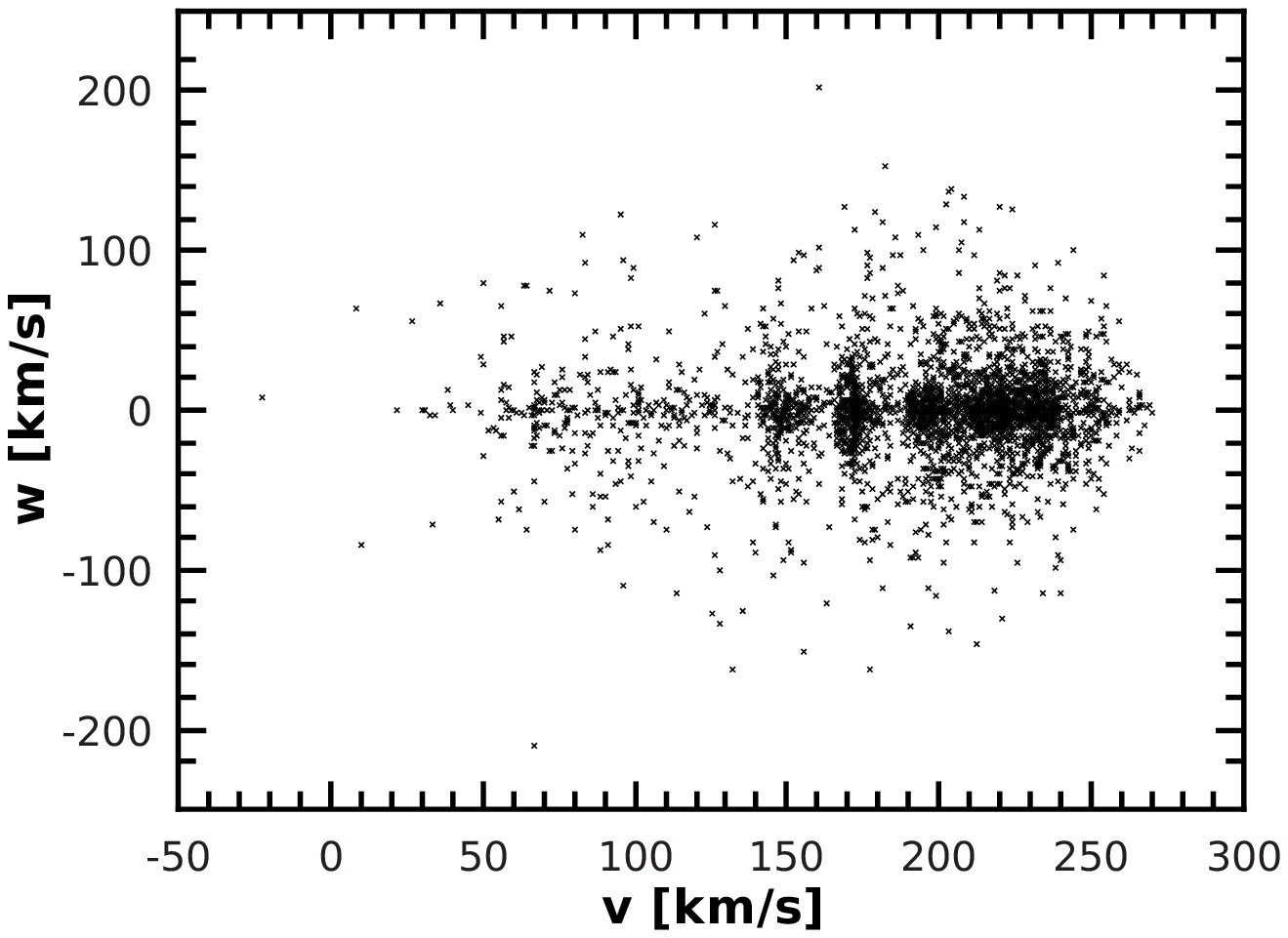}
  \caption{Evolution of a model SC in velocity space.  Left panels
    show the velocity distributions after $270$~Myr ($\approx$ one
    revolution round the Galaxy).  Right panels show the
    distributions at the end of the simulation ($10$~Gyr).  From top
    to bottom we show $W$ against $R$, $W$ against $U$ and finally $W$
    against $V$.  Top left panel shows clearly the quantized nature of
    the $W$-velocities which lead to the so-called Christmas-tree
    effect. At the lower right panel we see the 'azimuthal fish'.}
  \label{fig:fish-1}
\end{figure*}

The shape of the $z$-distribution looks like a Christmas tree
(this effect was seen in our simulations shortly before Christmas
2009).  The origin of this shape can be explained by considering the
dynamical processes during the SC evolution.  The star cluster is
orbiting on a circular orbit around the Galactic Centre.  After gas
expulsion, some of the stars acquire higher expanding velocities and
form the thick disc.  We show in Fig~\ref{fig:fish-1} in the top left
panel the $W$-velocity distribution as a function of the distance
between $8$ and $9$~kpc from the Galactic Centre after $270$~Myr
(i.e. about one orbital period).  The velocities show discrete layers.
A similar behaviour is also described in \citet{kuepper10} where the
tidal tails of SCs show density enhancements with regular spacings
stemming from the turn-around points of the escaped stars on
epicycles.  We suspect that our Christmas tree has a similar
explanation and is due to the turn-around points of stars on their
$z$-oscillations ($z$-epicycle).  \citet{kuepper10} report that their
star clusters have to be sufficiently long lived to show this
behaviour in their tidal tails, while we see this effect very early
in the evolution of our models.  The reason for this difference is
quite simple.  In the simulations of \citet{kuepper10} the star
clusters are in virial equilibrium and loose stars through tidal and
two-body effects quite slowly. Therefore it needs a long time for the
tidal tails to form and evolve. Thus, in the \citet{kuepper10} case
the epicyclic motions of stars evaporating from star clusters are
nearly coherent or in-phase since the stars leave the cluster with
very similar and small velocities.  The popping clusters, on the other
hand,  expel stars with a larger range of velocities within a short
time-span.

The middle and lower left panel of Fig.~\ref{fig:fish-1} show that we
only see this phenomenon in the $W$-velocities, while $U$ (radial
velocity) and $V$ (azimuthal velocity) show a simple bulk motion
(except for a few stars which lag behind).   This special appearance
of the Christmas tree gets later erased and transformed into a smooth
distribution as described in Sect.~\ref{sec:space}.  Also the
velocities in $W$ do not show the quantized behaviour any longer as we
can see in the top right panel of Fig.~\ref{fig:fish-1}.  The
transient tree feature more or less disappears when the stars which
form the leading arm of the tidal tails wrap all around the whole
orbit and start to overlap with the trailing stars filling the gaps in
velocity space.  A deeper theoretical explanation for this astonishing
feature is not part of this paper and will be dealt with in a future
publication.

In the right panels of Fig.~\ref{fig:fish-1} we show the particle
velocities after $10$~Gyr of simulations.  While we clearly see that
the quantized nature in the $W$-component has washed out (see above) a
new feature has appeared.  In all our simulations we see over- and
under-densities in the azimuthal $V$-velocities.  This quantization is
not as sharp as formerly seen in $W$ but still clearly visible.  We
call this feature azimuthal fish, because of its form.  We believe
that this quantization is the direct counterpart of the density
enhancements in tidal tails which are reported by \citet{kuepper10}.
The only difference between their and our models is that they
investigate slowly evolving globular clusters (GCs), which loose their
mass slowly over long time-intervals, while our models have dissolved
rapidly and the stars spread out much faster and wider than seen in
\citet{kuepper10}.  We will investigate this problem further and
report about it in a follow up paper.

\section{Conclusions}
\label{sec:conc}

We compute models of dissolving embedded star clusters on circular
orbits around the MW.  The aim of our study is to investigate if SCs 
of different mass, SFE and gas-expulsion time can support not only the
distribution of stars and their velocities in the thin disc of the MW
but also in the thick disc.

We also co-added the results of our simulations to show the effect of
a whole population of SCs, following a standard ICMF, to the
distribution of stars in the discs of the MW.

We can summarize the results of our models in the following points:
\begin{itemize}
\item The stars of dissolved SCs form density distributions in the
  $z$-direction which can be best fitted by exponential profiles.
\item While low-mass SCs (masses of the order of $10^{4}$ and
  $10^{5}$~M$_{\odot}$) contribute only to a component similar to the
  thin disc of the MW, high mass clusters(masses $>
  10^{5}$~M$_{\odot}$) show distributions which can be described best
  by two exponential profiles and therefore their stars contribute
  into two components similar to the thin and thick disc of the MW.
\item The velocity dispersion of the stars distributed into the thin
  component show a much lower velocity dispersion than the actual
  dispersion of the thin disc of the MW.  This may be due to
  the fact that we do not take any other mechanisms to enhance the
  velocity dispersion into account (e.g.\ spiral arms, giant molecular
  clouds, other SCs, ...)
\item The thick component of our model shows a velocity dispersion
  which is higher than observed in the MW thick disc.  We therefore
  conclude that inside the MW disc SCs with masses comparable to
  $10^{7}$~M$_{\odot}$ might never have formed.  The thick component
  of our $10^{6}$~M$_{\odot}$ model has enough velocity dispersion to
  explain the thick disc.  This is in nice agreement with the
  analytical results of \citet{Kroupa2002}.
\end{itemize}

We need to discuss these findings a bit further.  On the first glance
it seems to be rather odd that stars of one dissolved star cluster
should spread into two distinct distributions with two distinct
velocity dispersions.  The density distribution of all our dissolved
clusters exhibits a very distinct profile showing a 'peak' around
small $z$-distances and a kind of 'envelope' of large $z$-distances.
The transition between these two parts is rather continuous and not
sharp.  We see that for low-mass clusters the 'peak' is more
pronounced than the 'envelope' and therefore they can be fitted best
by a single exponential profile.  Also the stars in the 'envelope' do
not extend to $z$-heights comparable to the thick disc.  For high-mass
clusters we see a very substantial 'envelope', reaching out to high
$z$-distances, and therefore are better described by two exponentials.
The same is true for the distribution of $W$-velocities.  The
distribution is similar in shape to the distribution of velocities
found in stars in the solar neighbourhood.  It is neither a single nor
a double Gaussian but rather a continuous overlay of many Gaussians.
Nevertheless, the usual observational procedure to analyse these
distributions is to fit two Gaussians to the probability plot and
assigning two velocity dispersions to it - one for the thin disc
component and one for the thick disc component.  So we followed the
same method.  Again for low-mass clusters there are no stars with high
velocities and therefore we only assign one velocity dispersions to
their distributions.

Another issue, which needs to be discussed, is the constant analytical
potential used in this study.  It is clear that in the distant past
the disc(s) of the MW were much smaller in mass and maybe in size.  A
'popping' star cluster would spread its stars to much larger
$z$-heights much more easily. So it would rather help our scenario to
explain the thick disc. Furthermore the adiabatic heating of stars due
to the growth in mass of the MW will enhance the velocity dispersion
of the stars and bring them onto orbits with higher $z$-distances. 

Besides, it is well known that the thick disc and thin disc components
have a different chemical history. The ratios of $\alpha$-(O,Mg,Si,Ca,
and Ti) elements, for example, are higher for stars in the thick disc
than for stars of the thin disc at given metallicity
\citep{bens07,koba11}. 
This indicates a short star formation time-scale in which enrichment
is dominated by Type II supernovae (SNe II) \citep{alves10}.
Supernovae events guide the gas expulsion process, that together with
all the other scattering mechanisms, might also help to move stars
from the 'peak' structure into the 'envelope' structure. This would
explain the problem that we do not see stars with thick disc chemical
properties amongst thin disc stars and, therefore, why we see
different populations of stars with different chemical properties in
the thin and thick disc of our MW now.  Another possibility would be
if the MW saw an early epoch of star formation, where the formation of
low-mass clusters was suppressed {  \citep[in comparison
  with the evidence for this from top-heavy galaxy-wide initial mass
  functions of stars in young galaxies][]{weid10}.}

We therefore are able to show that one does not need any merger
scenarios to explain the structure of the MW disc.  The simple
assumption that all stars form in SCs is enough to explain the thin as
well as the thick disc of our MW, thus confirming the analytical
calculations by \cite{Kroupa2002} rather nicely.  These models are
applied by \cite{Kroupa2002} as a possible explanation not only of the
thick disk, but also of the observed but hitherto not understood
secular heating of the thin disk.  Here the notion raised is that the
observed thickening of the MW thin disk with age may be due to a
falling star-formation rate (SFR) since the past 10~Gyr, whereby a
small SFR implies the formation of SCs with small masses
\citep{weid04}.

While conducting our experiments we detected two new effects within
the distributions of the stars.  After about one revolution around the
Galaxy the stars have spread in the $z$-direction in to a form that
has striking similarities to a Christmas-tree.  As the discovery was
made shortly before Christmas we called the effect officially
Christmas-tree distribution.  At the same time the distribution of
$W$-velocities seems to have only certain discrete values.  This
feature is transient and disappears after several revolutions about the
MW.  We suspect that this is the case when the leading and trailing
arms wrap completely around the Galaxy and overlap at the former SC
position.  Then the velocities of the leading arm fall into the gaps
of the velocity distribution of the trailing stars.  The reason to see
a peculiar distribution like the Christmas-tree in the first place is
still not fully investigated and will be dealt with in a follow-up
paper.  We suspect that we see a similar behaviour in $z$ as is
reported by \citet{kuepper10} along the tidal tails.

The second effect showed up towards the end of our simulations.  This
effect consists of peculiar over- and under-densities in the $V$
velocities which show a pattern that reminded us of a fish.  We
therefore called this effect the azimuthal fish.  Even though the
'tidal tails' of the dissolved star clusters are well mixed towards
the end of the simulations so that we can not see any density
variations in the positions, we believe that our fish pattern is
directly related to the slowly evolving tidal tails of GCs and their
density variations.  As the quantized velocities associated with the
Christmas-tree we see now the left-over velocity signatures of a
similar effect in azimuthal direction., while the positional
counterpart is not visible due to the fast evolution of our models.
Also this effect will be investigated further and reported about in a
following publication.

As a final comment we state again that our models are able to
reproduce the structural features of the MW.  Our models add therefore
another possible explanation of how the discs of the MW and similar
galaxies may have acquired their properties.  Which theory will be
right - future observation might tell or even show that we might deal
with a superposition of all theories at the same time. \\

{\bf Acknowledgments:}
PA is supported through a Chilean CONICYT grant and a Deutscher
Akademischer Austauschdienst (DAAD) grant.  MF acknowledges support
through FONDECYT project no.\ 1095092.

\label{lastpage}

\end{document}